\documentclass[fleqn,usenatbib]{mnras}
\usepackage{newtxtext,newtxmath}
\usepackage[T1]{fontenc}
\DeclareRobustCommand{\VAN}[3]{#2}
\let\VANthebibliography\thebibliography
\def\thebibliography{\DeclareRobustCommand{\VAN}[3]{##3}\VANthebibliography}
\usepackage{graphicx}	
\usepackage{amsmath}	
\usepackage{soul}
\usepackage{comment}
\usepackage{caption}
\usepackage{subcaption}
\usepackage{appendix}
\usepackage{xcolor}
\usepackage{ulem,tikz}
\newcommand{\Cristiano}[1]{{\textcolor{orange}{\textt{Cristiano:#1}}}}
\newcommand{\KS}[1]{{\textcolor{purple}{\textt{Kendall:#1}}}}
\definecolor{tiffany}{RGB}{79, 166, 158}
\definecolor{bludodger}{RGB}{30, 144, 255}
\definecolor{applegreen}{rgb}{0.55, 0.71, 0.0}

\newcommand{\msc}[1]{\textcolor{tiffany}{[Mario - #1]}}
\setstcolor{red}

\title[Cosmic Rate of PISNe]{The cosmic rate of Pair-Instability Supernovae}

\author[F. Gabrielli et al.]{
\parbox{\linewidth}{Francesco Gabrielli$^{1,2}$\thanks{E-mail: fgabriel@sissa.it}, Andrea Lapi$^{1,2,3,4}$, Lumen Boco$^{1,2,3}$, Cristiano Ugolini$^{1,8}$, Guglielmo Costa$^{7}$, Cecilia Sgalletta$^{1,5}$, Kendall Shepherd$^{1,6}$, Ugo N. Di Carlo$^{1}$, Alessandro Bressan$^{1}$, Marco Limongi$^{8,9,10}$, Mario Spera$^{1,2,8}$}\\ \\
$^{1}$SISSA, via Bonomea 265, 34136 Trieste, Italy\\
$^{2}$National Institute for Nuclear Physics - INFN, Sezione di Trieste, I-34127 Trieste, Italy\\
$^{3}$Institute for fundamental physics of the Universe - IFPU, Via Beirut 2, 34014 Trieste, Italy\\
$^{4}$Istituto di Radioastronomia - INAF/IRA, Via Piero Gobetti 101, 40129 Bologna, Italy\\
$^{5}$INFN-Padova, Via Marzolo 8, I–35131 Padova, Italy\\
$^{6}$Osservatorio Astronomico di Padova-INAF, Vicolo dell’Osservatorio 5, I-35122 Padova, Italy\\
$^{7}$Univ Lyon, Univ Lyon1, Ens de Lyon, CNRS, Centre de Recherche Astrophysique de Lyon UMR5574, F-69230 Saint-Genis-Laval, France\\
$^{8}$ Istituto Nazionale di Astrofisica - Osservatorio Astronomico di Roma, Via Frascati 33, I-00040, Monteporzio Catone, Italy\\
$^{9}$ Kavli Institute for the Physics and Mathematics of the Universe, Todai Institutes for Advanced Study, University of Tokyo, Kashiwa, 277-8583, Japan\\
$^{10}$ INFN. Sezione di Perugia, via A. Pascoli s/n, I-06125 Perugia, Italy}

\date{Accepted XXX. Received YYY; in original form ZZZ}

\pubyear{XXX}

\begin{document}

\label{firstpage}
\pagerange{\pageref{firstpage}--\pageref{lastpage}}
\maketitle


\begin{abstract}
Pair-instability supernovae (PISNe) have crucial implications for many astrophysical topics, including the search for very massive stars, the black hole mass spectrum, and galaxy chemical enrichment. To this end, we need to understand where PISNe are across cosmic time, and what are their favourable galactic environments. We present a new determination of the PISN rate as a function of redshift, obtained by combining up-to-date stellar evolution tracks from the \texttt{PARSEC} and \texttt{FRANEC} codes, with an up-to-date semi-empirical determination of the star formation rate and metallicity evolution of star-forming galaxies throughout cosmic history. We find the PISN rate to exhibit a huge dependence on the model assumptions, including the criterion to identify stars unstable to pair production, and the upper limit of the stellar initial mass function. Remarkably, the interplay between the maximum metallicity at which stars explode as PISNe, and the dispersion of the galaxy metallicity distribution, dominates the uncertainties, causing a $\sim$ seven-orders-of-magnitude PISN rate range. Furthermore, we show a comparison with the core-collapse supernova rate, and study the properties of the favourable PISN host galaxies. According to our results, the main contribution to the PISN rate comes from metallicities between $\sim 10^{-3}$ and $10^{-2}$, against the common assumption that views very-low-metallicity, Population III stars as exclusive or dominant PISN progenitors. The strong dependencies we find offer the opportunity to constrain stellar and galaxy evolution models based on possible future (or the lack of) PISN observations.
\end{abstract}

\begin{keywords}
galaxies: general -- supernovae: general -- stars: evolution
\end{keywords}


\section{Introduction}

Pair-instability supernovae (PISNe) are explosions that develop inside the cores of very massive stars (VMSs) at the end of their evolution, leading to the complete disruption of the progenitor. 
The physical mechanism behind PISNe has been well understood ever since its discovery (\citealt{Fowler_1964,Bisnovatyi_Kogan_1967,Rakavy_1967,Barkat_1967,Fraley_1968}). At the end of C burning in the core, where temperatures approach $10^9\:K$ and densities are greater than $\sim 100\:g\:cm^{-3}$, photons become energetic enough to create electron-positron pairs. The pair-production process removes radiation pressure, which counteracts the gravitational pull from the inner layers, and lowers the adiabatic index $\Gamma$ below 4/3. As a result, the star becomes unstable and begins to collapse in a runaway fashion. The onset of explosive O and Si burning releases energies of $\sim 10^{52}-10^{53}$ ergs (\citealt{Heger_2002}), high enough to eject all the star's material, without leaving any remnant behind. The explosion produces high amounts of $^{56}$Ni, up to $\gtrsim 50\:M_{\rm \odot}$ (\citealt{Heger_2002}). Its radioactive decay is held responsible for luminosities up to $10^2$ times those of typical core-collapse supernovae (CCSNe, e.g. \citealt{Scannapieco_2003,Kasen_2011,Dessart_2012,Whalen_2013,Kozyreva_2014,Kozyreva_2014_1,Jerkstrand_2015,Smidt_2015,Kozyreva_2016,Gilmer_2017,Hartwig_2018,Chatzopoulos_2019}). Despite PISNe being so luminous, and the several hundreds of CCSN observations achieved so far (e.g. \citealt{Cooke_2012,Yaron_2012,Gal-Yam_2013,Guillochon_2017}), no PISN has ever been confidently discovered. Several candidate detections have been reported, including super-luminous supernovae (SLSNe), but none has been confirmed as a PISN (\citealt{Woosley_2007,Gal_Yam_2009,Quimby_2011,Gal_Yam_2012,Cooke_2012,Kozyreva_2015,Lunnan_2016,Kozyreva_2018,Mazzali_2019,Nicholl_2020,Gomez_2019,Schulze_2023}). 

PISNe are predicted to occur in stars with masses on the zero-age main sequence (ZAMS) in the range $140\lesssim M_{\rm ZAMS}/M_{\rm \odot}\lesssim 260$, and metallicities ($Z$) below some threshold (\citealt{Heger_2002,Heger_2003}). Stars undergo pair instability starting from $M_{\rm ZAMS}\sim 100\:M_{\rm \odot}$, but below $\sim$140 $M_{\rm \odot}$ they experience a series of pulsations, accompanied by the ejection of the most external layers, in a pulsational pair-instability supernova (PPISN, \citealt{Heger_2002,Heger_2003}). In this case, the core stays mostly intact, and the star finally collapses into a black hole (BH). Above $\sim 140\:M_{\rm \odot}$, the first pulsation is so energetic that it completely disrupts the star. For masses higher than $\sim 260\:M_{\rm \odot}$, the star directly collapses into an intermediate-mass BH (\citealt{Heger_2002,Heger_2003}). If $Z$ is too high, mass loss due to stellar winds prevents the star from forming cores massive enough to become unstable (\citealt{Heger_2003}). The maximum metallicity at which stars explode as PISNe is uncertain. Stellar evolution simulations succeed in producing PISNe up to some fraction of the solar metallicity, generally below $\sim 0.5\:Z_{\rm \odot}$ (\citealt{Langer_2007,Kozyreva_2014_1,Spera_2017,Langer_2012,Costa_2021,Higgins_2021,Sabhahit_2023,Martinet_2023}).

Stellar evolution codes simulate the evolution of stars from the ZAMS throughout the nuclear burning stages, providing a link between $M_{\rm ZAMS}$ and the final core mass, at the pre-SN stage. In order to identify stars undergoing pair instability, it is common to adopt a criterion based on the final mass of the core. As described above, the physical processes behind the onset of instability are much more complex. Nonetheless, the core mass criterion represents a good and useful proxy. By assuming that stars with He or CO core mass in a certain range end their life as PISN, it is possible to obtain a range of PISN progenitor masses. 
Depending on the details of the adopted stellar evolution code, the $M_{\rm ZAMS}$ range where PISNe occur can fluctuate (e.g. \citealt{Heger_2002,Heger_2003,Langer_2007,Takahashi_2015,Spera_2017,Woosley_2017,Marchant_2019,Iorio_2022,Tanikawa_2022}). Moreover, detailed stellar evolution calculations show that the PISN range tends to shift to higher $M_{\rm ZAMS}$ at increasing metallicity (e.g. \citealt{Heger_2002,Heger_2003,Spera_2017}). 

Since PISNe are expected to occur only in low-metallicity stars, pristine, very-low-metallicity Population III stars (Pop III) have been traditionally considered as main PISN progenitors (e.g. \citealt{Ober_1983,El_Eid_1983,Baraffe_2001,Umeda_2002,Heger_2002,Scannapieco_2005,Wise_2005,Langer_2007,Kasen_2011,Pan_2012,Dessart_2012,Yoon_2012,de_Souza_2013,de_Souza_2014,Whalen_2013,Whalen_2014,Smidt_2015,Magg_2016,Regos_2020,Venditti_2023,Bovill_2024,Wiggins_2024}). 
However, the fact that stellar evolution simulations allow for PISNe up to $\sim 0.5\:Z_{\rm \odot}$, suggests that also higher-$Z$, Population II/I (Pop II/I) stars might give a significant contribution to the PISN rate.

Comprehending the physics behind and the occurrence of PISNe, particularly in relation to their surrounding environments, holds a myriad of astrophysical implications. 

For instance, the study of PISNe is strongly linked to the debate on the upper limit of the stellar initial mass function (IMF). Stars with masses consistent with $\gtrsim 200-300\:M_{\rm \odot}$ have indeed been observed in local galaxies (\citealt{Crowther_2010,Evans_2010,Walborn_2014,Crowther_2014,Schneider_2018,Crowther_2019,Bestenlehner_2020,Kalari_2022,Brands_2022}), challenging the previous consensus that placed the maximum stellar mass at $\sim$150 $M_{\rm \odot}$ (\citealt{Vink_2015}).
The existence of VMSs is also supported by chemical abundances studies (e.g. \citealt{Romano_2017,Romano_2017_erratum,Romano_2020,Goswami_2021,Goswami_2022}). The observation of massive stellar BH binaries in gravitational waves (GWs) offers a new opportunity to investigate VMSs, even though all confident detections achieved so far do not necessarily require such massive progenitors (e.g. \citealt{Abbott_2016,Abbott_2020,Spera_2015,Spera_2017,Vink_2021}). Due to the very-high masses of PISN progenitors, the location of the IMF upper limit can critically determine the PISN rate.

Moreover, PISNe can help to shed light on many uncertain aspects of galaxy evolution. Indeed, PISN occurrence strongly depends on the properties of the galactic environments in which they take place. Therefore, their study requires a determination of the $Z$-dependent SFR density (SFRD) across cosmic history. This quantity defines the amount of mass available to form stars in the Universe, per unit time, comoving volume and metallicity. To estimate it, two main approaches are usually adopted, relying either on cosmological simulations (e.g. \citealt{Mapelli_2017,O_Shaughnessy_2016,Lamberts_2018,Mapelli_Giacobbo_2018,Artale_2019}), or on empirical prescriptions for the SFRD and galaxy metallicity distribution, derived from observations (e.g. \citealt{Belczynski_2016_1,Lamberts_2016,Cao_2017,Elbert_2017,Li_2018,Boco_2019,Boco_2021,Chruslinska_2019,Neijssel_2019,Santoliquido_2021}). Many uncertainties still exist around this subject (see e.g. \citealt{Chruslinska_2019,Neijssel_2019,Boco_2021,Santoliquido_2021} for a comprehensive overview). In particular, the metallicity distribution of galaxies is still considerably unknown. Moreover, the low-mass end of the galaxy stellar mass functions (GSMFs), describing the mass distribution of galaxies, is still poorly constrained, since low-mass galaxies are very faint and thus difficult to observe. Specifically, it is not clear whether the GSMF slope in this mass range is constant, or redshift dependent (see e.g. \citealt{Navarro_Carrera_2023}, where the latter case is supported by JWST data at redshifts $4\lesssim z\lesssim 8$).

PISNe are also commonly invoked in studies about the chemical enrichment of the Universe (e.g. \citealt{Ricotti_2004,Matteucci_2005,Ballero_2006,Cherchneff_2009,Rollinde_2009,Cherchneff_2010}), and in particular to explain the chemical abundance patterns observed in the Milky Way and local galaxies (e.g. \citealt{Kojima_2021,Goswami_2021,Goswami_2022}). The detection of PISN descendants, i.e. stars with chemical abundances compatible with at least partial enrichment by a PISN, represents another avenue to find out about PISN occurrence, besides direct observation (e.g. \citealt{Heger_2002,Aoki_2014,Takahashi_2018_b,Salvadori_2019,Caffau_2022,Aguado_2023,Xing_2023,Koutsouridou_2023}). Hence, obtaining insights into the existence and rate of PISNe across cosmic time would have strong implications for numerous unresolved inquiries in both astrophysics and cosmology.

In this work, we compute the PISN rate as a function of redshift, by combining up-to-date stellar evolution tracks from the \texttt{PARSEC} and \texttt{FRANEC} codes, with an up-to-date semi-empirical determination of the $Z$-dependent SFRD across cosmic history. The aim is to study the dependence of the PISN rate on stellar and galactic prescriptions, an aspect which has not been explored extensively in the past. We also compute the ratio between PISN and CCSN rate, in order to provide a comparison with these observed transients. Finally, we explore the properties of the favourable PISN host galaxies. This work represents the premise to a follow-up work, where we will employ the theoretical framework presented here to study PISN observability with the \textit{James Webb Space Telescope} (JWST), and address the question why these transients have never been observed. These works offer the opportunity to put constraints on stellar and galaxy evolution models, via the comparison with possible future PISN observations, or with the absence of observations in the eventuality that they are never discovered.

The paper is structured as follows. In Section \ref{sec:model}, we present our theoretical framework. We start by describing our semi-empirical, $Z$-dependent SFRD determination, and how we use stellar evolution tracks to compute the number of PISNe produced per unit star forming mass. We also show the considered variations on stellar and galactic prescriptions, and how we finally compute the PISN rate. Our results are presented in Section \ref{sec:results}, and discussed in Section \ref{sec:discussion}, where we also show additional variations. We draw our conclusions in Section \ref{sec:conclusions}.

Throughout this work, we assume the flat $\Lambda$CDM cosmology from \citealt{Planck_2020}, with parameters $\Omega_M=0.32$, $\Omega_b=0.05$, $H_0=67\:km\:s^{-1}\:Mpc^{-1}$. A standard Kroupa IMF is adopted (\citealt{Kroupa_2001}), defined from 0.1 $M_{\rm \odot}$. The IMF upper limit is among the parameters we decide to vary. Following \citealt{Caffau_2010}, we use the value of $Z_{\rm \odot}=0.0153$ for the Solar metallicity, and $12+\log$(O/H)$_{\rm \odot}=8.76$ for the Solar oxygen abundance. 

\section{Methods}\label{sec:model}
We take into account how galaxies evolve throughout cosmic history by constructing a detailed semi-empirical determination of the $Z$-dependent SFRD, $d^3M_{\rm SFR}/dtdVd\log Z$, directly based on observations, following \citealt{Boco_2021}. This provides us with the amount of mass available for star formation at a certain redshift, per unit time, comoving volume and metallicity. In order to compute the number of PISNe produced per unit star forming mass, $dN_{\rm PISN}/dM_{\rm SFR}$, we make use of stellar evolution tracks computed with the \texttt{PARSEC} code (\citealt{Bressan_2012,Chen_2014,Tang_2014,Chen_2015,Costa_2019,Costa_2021,Nguyen_2022}). Finally, we compute the PISN rate density as a function of redshift by convolving these two quantities, according to the following formula:
\begin{equation}\label{eq:rate_z}
\frac{d^2N_{\rm PISN}}{dtdV}(z)=\int d\log Z\:\frac{d^3M_{\rm SFR}}{dtdVd\log Z}(Z,z)\times\frac{dN_{\rm PISN}}{dM_{\rm SFR}}(Z).
\end{equation}
We describe in detail how we compute each of these quantities in the following Sections. In order to account for the uncertainties in stellar and galaxy evolution models, we decide to follow a variational approach, considering alternative stellar evolution codes, and different values for the relevant stellar and galactic parameters.

\subsection{Galaxy evolution}\label{sec:gal_model}
We follow the semi-empirical approach presented in \citealt{Boco_2021} in order to compute the $Z$-dependent SFRD, with some updates informed by more recent works (\citealt{Chruslinska_2021,Popesso_2022}). This approach is built on galaxy observations up to $z=6$. First, we compute galaxy stellar mass functions (SMFs), providing a galaxy statistics based on their stellar mass, $\Phi(M_{\star})=d^2N/dVd\log M_{\rm \star}$. Here, $N$ indicates the number of galaxies, $M_{\rm \star}$ their stellar mass, and $V$ the comoving volume. We then convolve the SMFs with the galaxy main sequence (MS), $\psi(M_{\rm \star})$, relating the galaxy stellar mass to their SFR, $\psi$. We also implement a distribution of galaxies as a function of SFR, $dp/d\log \psi$, as indicated by observations (see below). In this way we are able to obtain SFR functions, i.e. a galaxy statistics based on SFR. For the galaxy metallicity, we define a log-normal distribution, $dp/d\log Z\:(Z,Z_{\rm FMR})$, with a constant dispersion $\sigma_{\rm Z}$. The mean value, $Z_{\rm FMR}$, is given by the Fundamental Metallicity Relation (FMR) linking galaxy mass, metallicity and SFR. We finally convolve all these factors according to the following formula, where we integrate over $M_{\rm \star}$ and $\psi$:
\begin{equation}\label{eq:CSFRD}
\begin{split}
\frac{d^3M_{\rm SFR}}{dtdVd\log Z}(Z,z) = & \int d\log M_{\rm \star}\frac{d^2N}{dVd\log M_{\rm \star}}(M_{\rm \star},z)\\
& \times \int d\log \psi\:\psi \frac{dp}{d\log \psi}(\psi,M_{\rm \star},z)\\
& \times \frac{dp}{d\log Z}(Z,Z_{\rm FMR}(M_{\rm \star},\psi))\\
\end{split}
\end{equation}
Each ingredient in Equation \ref{eq:CSFRD} is computed as described in the following. 

\subsubsection*{\textbf{SMF+MS}}

\begin{figure}
	\includegraphics[width=\columnwidth]{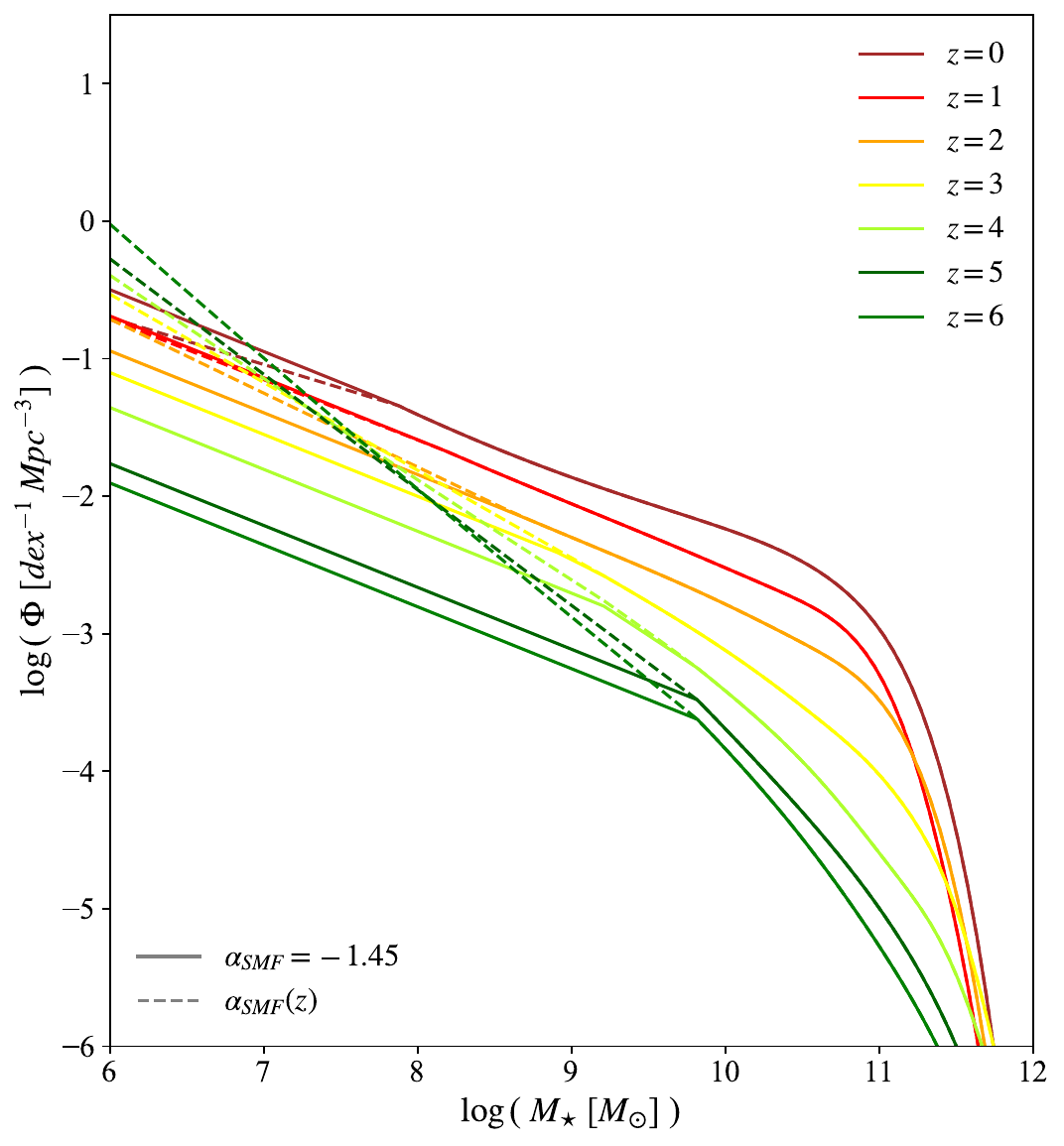}
    \caption{SMFs computed following \citealt{Chruslinska_2019}, at redshift $z=0$ to 6. We show both variations with $\alpha_{\rm SMF}=-1.45$ and $\alpha_{\rm SMF}(z)=-0.1z-1.34$, in solid and dashed lines respectively.}
    \label{fig:gsmf}
\end{figure}

We follow the work by \citealt{Chruslinska_2019} in order to compute the galactic SMFs. They perform a comprehensive determination by gathering several previous results from the literature, consisting in Schechter and/or double Schechter analytical fits to observational data. In particular, they consider discrete redshift bins, and in each of them they average among previous results in order to obtain the SMFs at the corresponding redshifts. The SMFs at arbitrary redshifts are computed by interpolating between these curves. In order to take into account the uncertainties around the low-mass end of the SMFs, they consider two variations, defining a constant low-mass end slope, $\alpha_{\rm SMF}=-1.45$, or prescribing a redshift dependence, $\alpha_{\rm SMF}(z)=-0.1z-1.34$. 
In Figure \ref{fig:gsmf} we show the SMFs we obtain by following the same procedure, for both $\alpha_{\rm SMF}$ cases (solid and dashed lines respectively), in the redshift range relevant to this work, $0\leq z\leq 6$. 

It is to be noted that \citealt{Chruslinska_2019} adopt a Kroupa IMF up to 100 $M_{\rm \odot}$, while we extend it up to 300 $M_{\rm \odot}$, as we will see later. This should not affect our results significantly, given that the Kroupa IMF predicts a relatively low number of massive stars between 100 and 300 $M_{\rm \odot}$. Therefore the SMFs are not expected to differ appreciably. 

\begin{figure}
	\includegraphics[width=\columnwidth]{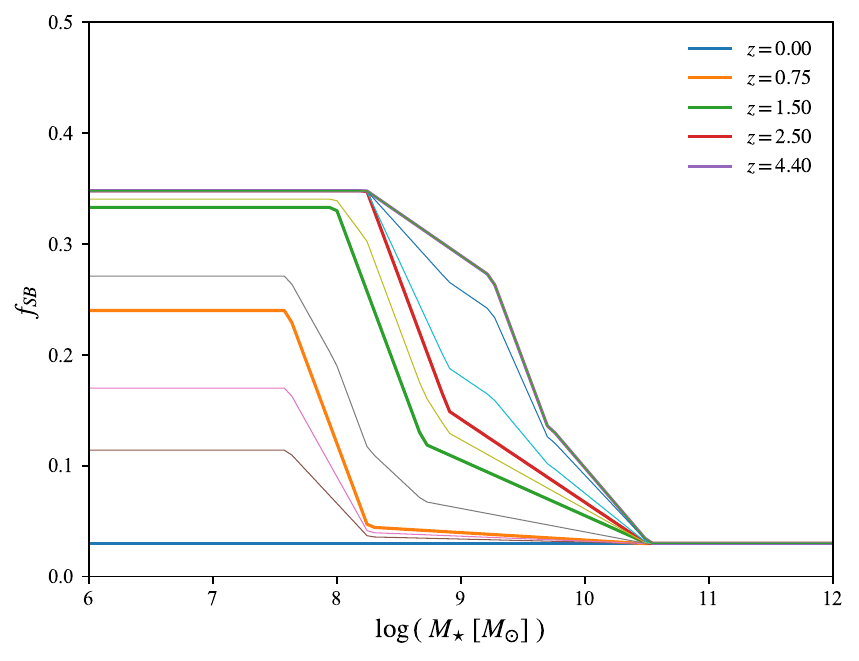}
    \caption{SB fractions $f_{\rm SB}$ as a function of $M_{\rm \star}$, for different $z$, computed following \citealt{Chruslinska_2021}. Thicker lines represent the $f_{\rm SB}$ computed for five initial redshifts (shown in the legend), while the $f_{\rm SB}$ at intermediate redshifts are computed by interpolation. At $z=0$, $f_{\rm SB}$ remains constant at 0.03, while above $z=4.4$ all curves saturate at 0.35, due to the dearth of observational data at those high redshifts.}
    \label{fig:sb}
\end{figure}

We define our galactic MS following \citealt{Popesso_2022}, which is the most complete determination up to date, taking into account all previous works in the literature and covering an unprecedented redshift and mass range, $0<z<6$ and $10^{8.5}$-$10^{11.5}$ $M_{\rm \odot}$. Additionally, we implement a double-Gaussian distribution in SFR, following \citealt{Sargent_2012}:
\begin{equation}\label{eq:ms_sb_scatter}
\begin{split}
    \frac{dp}{d\log \psi}(\psi,M_{\rm \star},z) &= \frac{f_{\rm MS}}{\sigma_{\rm MS}\sqrt{2\pi}}\:\exp\left[-\frac{\left(\log \psi-\langle \log \psi\rangle_{\rm MS}\right)^2}{2\sigma_{\rm MS}^2}\right] +\\
    & + \frac{f_{\rm SB}}{\sigma_{\rm SB}\sqrt{2\pi}}\:\exp\left[-\frac{\left(\log \psi-\langle \log \psi\rangle_{\rm SB}\right)^2}{2\sigma_{\rm SB}^2}\right],
\end{split}
\end{equation}
where $\langle \log \psi\rangle_{\rm MS}$ is the MS relation, and $\sigma_{\rm MS}=0.188$. As one can note by looking at the second term, we also consider the presence of starbursts (SBs), i.e. galaxies experiencing intense star formation, located in a separate region above the MS in the $\psi-M_{\rm \star}$ plane. The mean of the SB distribution is given by $\langle \log \psi\rangle_{\rm SB}=\langle \log \psi\rangle_{\rm MS}+0.59$, while $\sigma_{\rm SB}=0.243$. $f_{\rm MS}$ and $f_{\rm SB}$ are the fractions of MS and SB galaxies, respectively, such that $f_{\rm MS}+f_{\rm SB}=1$. Differently from \citealt{Boco_2021}, where a fixed SB fraction $f_{\rm SB}=0.03$ is assumed, we implement a dependence of $f_{\rm SB}$ on galaxy stellar mass and redshift, as done in \citealt{Chruslinska_2021}. In Figure \ref{fig:sb} we show the $f_{\rm SB}$ we compute following their work. This prescription enhances $f_{\rm SB}$ at low masses and increasing redshifts, bringing it from 0.03 up to 0.35. After $z=4.4$, $f_{\rm SB}$ saturates at values constant in $z$. All in all, the $M_{\rm \star}$ and $z$ dependencies enter in Equation \ref{eq:ms_sb_scatter} through $\langle \log \psi\rangle_{\rm MS}$, $\langle \log \psi\rangle_{\rm SB}$, $f_{\rm MS}$ and $f_{\rm SB}$.

\subsubsection*{\textbf{Z distribution}}\label{sec:Z_distr}
The FMR prescribes a correlation between galaxy stellar mass, SFR and metallicity, $Z_{\rm FMR}(M_{\rm \star},\psi)$. We define it following \citealt{Curti_2020}:
\begin{equation}
Z_{\rm FMR}(M_{\rm \star},\psi)=8.779-(0.31/2.1)\times\log\left(1+(M_{\rm \star}/M_0(\psi))^{-2.1}\right),
\end{equation}
where $M_0(\psi)=10^{10.11}\times\psi^{0.56}$.

Furthermore, we assume $Z$ to follow a log-normal distribution around the FMR,
\begin{equation}
\frac{dp}{d\log Z}(Z,Z_{\rm FMR}(M_{\rm \star},\psi))\propto \exp\left[-\frac{\left(\log Z-\log Z_{\rm FMR}(M_{\rm \star},\psi)\right)^2}{2\sigma_{\rm Z}^2}\right],
\end{equation}
We consider variations $\sigma_{\rm Z}=[0.15, 0.35, 0.70]$, to study the effect of this parameter on the PISN rate.\\

\begin{figure*}
    \includegraphics[width=0.8\textwidth]{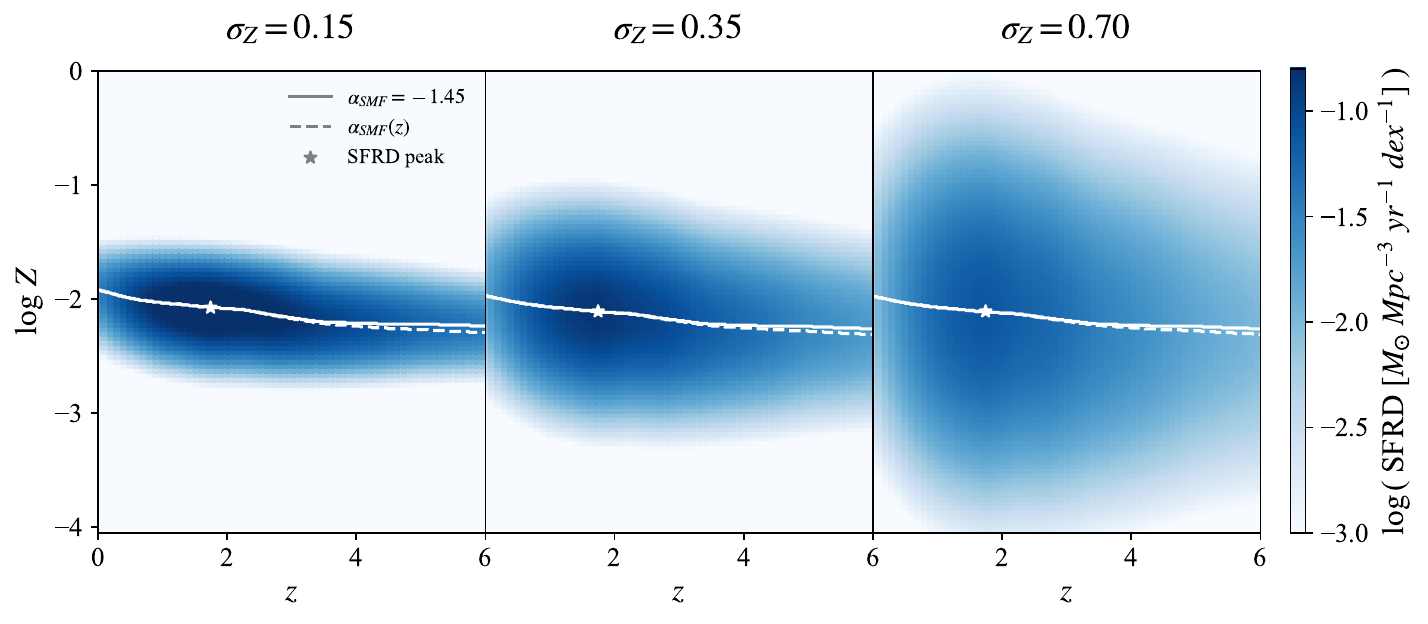}
    \caption{SFRD as a function of redshift and metallicity, for $\sigma_{\rm Z}=[0.15, 0.35, 0.70]$ (from left to right), and $\alpha_{\rm SMF}=-1.45$. The colorbar shows the SFRD values. White lines indicate the metallicity at which the SFRD peaks, as a function of redshift, for $\alpha_{\rm SMF}=-1.45$ (solid lines), and also $\alpha_{\rm SMF}=\alpha_{\rm SMF}(z)$ (dashed lines), for comparison. White stars indicate the redshift corresponding to the SFRD peak.}
    \label{fig:CSFRD_in_Z}
\end{figure*}

\begin{figure}
	\includegraphics[width=\columnwidth]{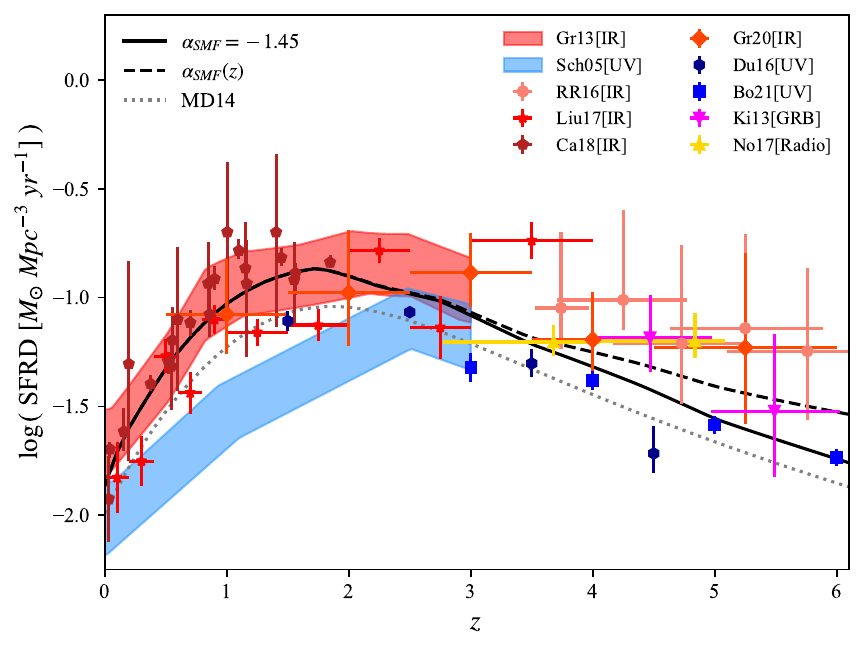}
    \caption{Cosmic SFR density as a function of redshift, for $\alpha_{\rm SMF}=-1.45$ and $\alpha_{\rm SMF}(z)$ variations (solid and dashed black lines respectively). The data points and bands show the observational determinations available at different redshifts, obtained from galaxy surveys in different bands (see the text for references). We also show the \citealt{Madau_2014} plot, corrected for our Kroupa IMF, for comparison (dotted grey line).}
    \label{fig:CSFRD}
\end{figure}

We report the final outcome of our galactic model in Figure \ref{fig:CSFRD_in_Z}, where we show the $Z$-dependent SFRD for all three variations on $\sigma_{\rm Z}$, fixing $\alpha_{\rm SMF}=-1.45$ (Equation \ref{eq:CSFRD}). We also show the metallicity corresponding to the SFRD peak, as a function of redshift, for both variations on $\alpha_{\rm SMF}$ (solid and dashed white lines), as well as the position of the overall SFRD peak (white star). In Figure \ref{fig:CSFRD} we instead report the SFRD as a function of redshift, obtained by integrating the previous quantity over $Z$, according to:
\begin{equation}
\frac{d^2M_{\rm SFR}}{dtdV}(z)=\int d\log Z \frac{d^3M_{\rm SFR}}{dtdVd\log Z}(Z,z).
\end{equation}
In order to show the agreement of our SFRD with observations, we also plot several empirical determinations from galaxy surveys in different bands (\citealt{Schiminovich_2005,Gruppioni_2013,Dunlop_2016,Rowan-Robinson_2016,Novak_2017,Casey_2018,Liu_2018,Gruppioni_2020,Bouwens_2021}), as well as the \citealt{Kistler_2009,Kistler_2013} data obtained from long gamma-ray burst observations.

Different ways to compute this quantity exist in the literature. Among empirical or semi-empirical approaches, one of the main alternatives is to rely on SFR functions (SFRFs), describing the number density of galaxies with a given SFR. They can be computed from galaxy UV and IR luminosity functions, using the conversion between luminosity and SFR. Moreover, it is possible to implement a $Z$-dependence on the SFRD using a Mass Metallicity Relation (MZR), linking galaxy mass and metallicity, instead of a FMR. Another common approach is to combine an existing SFRD determination with a standalone galaxy $Z$ distribution in redshift. We refer to \citealt{Boco_2021} for a comparison between these different methods. Finally, we stress that one of the main advantages of empirical determinations like ours, with respect to those relying on cosmological simulations, is that they are free from theoretical assumptions, being directly informed by observations.

\subsection{Stellar evolution}\label{sec:st_ev_model}
The second part of our method consists in computing the number of PISNe produced per unit star forming mass available, $dN_{\rm PISN}/dM_{\rm SFR}$. We do so by integrating an assumed IMF, $\phi(M)$, over the mass range of PISN progenitors, according to the following formula:
\begin{equation}\label{eq:stellar_part}
\frac{dN_{\rm PISN}}{dM_{\rm SFR}}(Z)=\frac{ \int_{\rm M_{\rm entry}}^{M_{\rm exit}} \phi(M) dM }{ \int_{\rm 0.1}^{M_{\rm up}} M \phi(M) dM},
\end{equation}
where $M_{\rm entry}$ and $M_{\rm exit}$ are the ZAMS masses for entering and exiting the PISN range, respectively. We adopt a Kroupa IMF \citep{Kroupa_2001} defined in the mass range [$0.1-M_{\rm up}$] $M_{\rm \odot}$, where the lower limit is the minimum stellar mass to ignite core H burning. We choose to vary $M_{\rm up}$ from 150 to 300 $M_{\rm \odot}$, in order to study the effects of different IMF upper limits on the PISN rate (see the Appendix \ref{appendix:IMF600} for an additional variation with $M_{\rm up}=600\:M_{\rm \odot}$). 

In order to compute $M_{\rm entry}$ and $M_{\rm exit}$, we adopt the criterion on the final core mass, according to which a star will develop pair instability leading to a PISN explosion if its He or CO core mass ($M_{\rm He/CO}$) in the pre-SN stage lies in a certain range. Then we employ stellar evolution tracks to link the initial, ZAMS mass to the final core mass, and obtain a $M_{\rm ZAMS}$ range for PISN progenitors. We gather several works studying the evolution of VMSs from the literature, where the authors consider a range of He and/or CO core masses for which stars end their life as PISN (\citealt{Heger_2002,Heger_2003,Langer_2007,Chatzopoulos_2012,Takahashi_2015,Spera_2016,Belczynski_2016,Spera_2017,Woosley_2017,Eldridge_2018,Marchant_2019,Farmer_2019,Stevenson_2019,Kinugawa_2020,Belczynski_2020,Marchant_2020,du_Buisson_2020,Tanikawa_2021,Briel_2022_1,Woosley_2021,Olejak_2022,Briel_2023,Tanikawa_2022}). Specifically, we decide to adopt the criterion on $M_{\rm CO}$, instead of $M_{\rm He}$, motivated by the fact that it is in the CO core that PI sets in. Moreover, varying C burning reaction rate, and consequently the C/O ratio at He core depletion, impacts significantly on the position of the star in the core density-temperature plane, i.e. on the onset of PI (e.g. \citealt{Farmer_2019}). In any case, as shown in Appendix \ref{appendix:He}, varying between CO and He core criterion does not significantly affect our results. Due to the uncertainties existing on these ranges, we consider an optimistic and pessimistic variation, featuring the widest and shortest intervals based on the literature, i.e. $M_{\rm CO}\in[60-105]$ $M_{\rm \odot}$ and $M_{\rm CO}\in[45-120]$ $M_{\rm \odot}$, respectively. We also consider an intermediate case with $M_{\rm CO}\in[55-110]$ $M_{\rm \odot}$.

We consider two sets of stellar evolution tracks computed with the \texttt{PARSEC} code, that follows the evolution of single stars from the pre-main sequence up to the most advanced burning phases. The first set (\citealt{Bressan_2012,Chen_2014,Tang_2014,Chen_2015}) was used in \citealt{Spera_2017}, implemented in the binary population synthesis code \texttt{SEVN}, and has been employed in several works in the last years. We will address it as \texttt{PARSEC-I}. The second and more recent set, which we will call \texttt{PARSEC-II} (\citealt{Costa_2019,Costa_2021}), was implemented in the recently published version of \texttt{SEVN} (\citealt{Iorio_2022}).

In \texttt{PARSEC-II}, the nuclear reactions and elemental mixing are coupled and solved at the same time in a diffusive scheme (\citealt{Marigo_2013, Costa_2019}). Additionally, \texttt{PARSEC-II} has updated prescriptions for the mass loss of massive stars, including Wolf-Rayet (WR) type wind (\citealt{Sander_2019}, see \citealt{Costa_2021} for more details). Moreover, these new models include an updated equation of state (EOS), also accounting for the effects of e$^-$e$^+$ pair creation.

In order to see the effects of varying stellar evolution code, we also consider the results obtained with the \texttt{FRANEC} stellar evolution code (\citealt{Chieffi_2013,Limongi_2018}). Comparing the results obtained using different sets of evolutionary tracks, especially when they are calculated by different groups that use different codes, is not a trivial procedure because, besides differences already arising from the different numerical procedures adopted, there are also variations due to different assumptions in the model calculations, and even in the description of physical complex phenomena that still require a calibration against observations. For example, different assumptions could refer to the criterion (basically Schwarzschild or Ledoux) adopted to establish whether a region is stable or not to convective motions. Instead, the calibration process may refer to the calibration against the Solar model to fix the value of the Mixing Length Theory (MLT), or to the size of the overshooting adopted in the convective core or at the bottom of the convective envelope. It will be impossible to trace back all these differences (and their impact on the results), and we will list below only the major differences that we believe must be considered when performing a sound comparison, leaving the details to the individual papers that describe the model calculations.

\texttt{PARSEC} models use the \citealt{Caffau_2010} solar partition of heavy elements (with $Z_{\odot}=0.0153$), even at very-low metallicity, while \texttt{FRANEC} adopts the \citealt{Asplund_2009} solar partition (with $Z_{\odot}=0.0134$), and consider $\alpha$-element enhancement at low metallicity (see e.g. \citealt{Limongi_2018}). This small discrepancy can cause models to evolve at slightly different luminosities already on the main sequence (\citealt{Sibony_2024}). Furthermore, since the metallicity of the Sun is used as a reference in the scaling law of the mass-loss process with metallicity, this small difference can induce differences in the mass-loss rates at different absolute metallicities, even adopting the same mass-loss prescriptions.

\texttt{PARSEC-I} provides tracks for 12 different metallicities from $1\times 10^{-4}$ to $3\times 10^{-2}$, while \texttt{PARSEC-II} considers 13 metallicities from $1\times 10^{-4}$ to $4\times 10^{-2}$. The \texttt{FRANEC} tracks are instead available for 4 metallicities, from $3\times 10^{-5}$ to $\sim 1.35\times 10^{-2}$. We linearly interpolate to obtain stellar tracks at metallicities in between, following \citealt{Iorio_2022}. Moreover, while the \texttt{PARSEC-II} tracks are computed for masses up to $600\:M_{\rm \odot}$, the \texttt{PARSEC-I} and \texttt{FRANEC} ones extend up to $\gtrsim$300 and 120 $M_{\rm \odot}$ respectively. We linearly extrapolate at higher masses. We caution that the actual trend of the stellar tracks at these high masses might deviate quite significantly from this approximation, which should be taken into account in examining our results.

For the convective stability, \texttt{PARSEC} models adopt the Schwarzschild criterion, while \texttt{FRANEC} models adopt the Ledoux one. The latter is more restrictive than the former and, if we consider that the core overshooting prescriptions are slightly different between \texttt{PARSEC} and \texttt{FRANEC}, and that \texttt{PARSEC} also accounts for overshooting from the bottom of the convective envelope, it is clear the the evolution in the low-mass range of massive stars may be different (but given the high non-linearity of the solutions, it is not easy to trace back the nature of the differences).

On the other hand, the evolution of more massive stars is strongly regulated by mass-loss (\citealt{Smith_2014}), and the effect that may have the strongest impact on the results is the adopted prescription for the mass-loss rates in the different evolutionary phases. Concerning massive stars, we remind that \texttt{PARSEC} includes radiative winds depending on the mass and evolutionary phase as described in \citealt{Chen_2015} and \citealt{Costa_2021}. In particular, in hot stars ($T_{\rm eff} \geq 10000\:K$), it uses the mass-loss prescriptions by \citealt{Vink_2000,Vink_2001}, including a surface iron abundance dependence; it also includes a dependence on the Eddington ratio (\citealt{Grafener_2008,Vink_2011}), which becomes important for the most luminous stars. For WR stars, i.e. when $T_{\rm eff} \geq 20000\:K$ and the hydrogen surface abundance is less than 0.3 in mass fraction, \texttt{PARSEC-II} uses the luminosity-dependent prescription for the mass loss from \citealt{Sander_2019}, while \texttt{PARSEC-I} uses a combination of literature mass-loss rates. Finally both \texttt{PARSEC-I} and \texttt{II} use the prescription by \citealt{deJager_1988} for cold massive stars (i.e. red super giants, RSGs).

In \texttt{FRANEC} models, the Eddington ratio is checked within each structure, and if a region where it is larger than unity is found, this region and the overlying layers are removed from the star. Furthermore, for RSGs a dust-driven mass-loss rate is used (\citealt{vanLoon_2005}). Both of these latter recipes for the mass-loss rates may cause strong differences in the evolution of the most massive stars, even for those with initial mass as low as $M_{\rm ZAMS}=15\:M_{\odot}$ or $20\:M_{\odot}$ at near-solar metallicity.

All stellar tracks employed here are calculated without rotation. The main effects of including rotation would be increased mass loss, and bigger cores due to enhanced chemical mixing. Overall, we expect the $M_{\rm ZAMS}$ ranges of PISN progenitors to shift to lower masses (see also the recent work by \citealt{Winch_2024}), resulting in higher PISN rates. It would be interesting to evaluate the entity of these effects by employing evolution tracks of rotating stars, which we do not explore here.

It is important to note that, as one can see, in this work we restrict to metallicities down to $10^{-4}$, and redshifts up to $z=6$, effectively only considering Pop II/I stars as PISN progenitors. This is motivated by the fact that huge uncertainties still exist around quantities such as the SFRD at very-high redshifts, $z>6$, and the Pop III IMF, which prevents from extending the study to Pop III stars in a comparably robust way. We discuss this issue in more depth in Section \ref{sec:pisn_popIII}.

Table \ref{tab:mzams_intervals} shows the $M_{\rm ZAMS}$ ranges obtained with each stellar evolution code, at some representative metallicities. We note that $M_{\rm up}$ cuts these ranges above its value. Masses greater than $300\:M_{\rm \odot}$ were fixed to $300\:M_{\rm \odot}$, since we only consider IMFs up to this value in this work (but see also the Appendix \ref{appendix:IMF600}). As one can see, these ranges can differ quite significantly from the canonical $[140-260]\:M_{\rm \odot}$. Moreover, they generally shift to higher masses at higher metallicities. Indeed, mass loss due to stellar winds is enhanced, thus the star must be initially more massive in order to produce a core satisfying the PISN criterion. Above a certain metallicity, which depends on the considered variation, the core cannot reach this mass threshold. We refer to the maximum metallicity for a star to explode as PISN, as $Z_{\rm max}$. As we will later show, $Z_{\rm max}$ turns out to be a crucial quantity in determining the PISN rate.

\begin{table*}
 \centering
 \captionof{table}{$M_{\rm ZAMS}$ ranges of PISN progenitors, for representative metallicities, obtained with the \texttt{PARSEC-I}, \texttt{PARSEC-II}, and \texttt{FRANEC} stellar evolution tracks. The ranges at metallicities in between the original ones from the stellar tracks are obtained via interpolation, as described in the text. All values are in solar units. Blanks indicate cases with no PISN, at $Z>Z_{\rm max}$. Masses $>300\:M_{\rm \odot}$ are fixed to $300\:M_{\rm \odot}$, i.e. the highest $M_{\rm up}$ considered in this work (see Appendix \ref{appendix:IMF600} for an additional variation with $M_{\rm up}=600\:M_{\rm \odot}$). The double intervals at $Z=1\times 10^{-4}$ of \texttt{PARSEC-II} are due to the non-monotonicity of the corresponding stellar track, which exits the PISN mass range and then re-enters again (see Figure 8 of \citealt{Iorio_2022}).}
 \begin{tabular}{|c|c|c|c|c|c|c|}
     \hline
     \raisebox{-2ex}{\textit{$M_{\rm CO}$}} \hspace{-0.6cm} \rotatebox[origin=c]{45}{$\Biggm\backslash$} \hspace{-0.3cm} \raisebox{1ex}{\textit{Z}} & $1\times 10^{-4}$ & $1\times 10^{-3}$ & $4\times 10^{-3}$ & $8\times 10^{-3}$ & $1\times 10^{-2}$ & $2\times 10^{-2}$ \\
     \hline
     \multicolumn{7}{|c|}{\texttt{PARSEC-I}}\\
     \hline
     45-120 & 108-257 & 109-300 & 158-300 & 178-222 & - & - \\
     \hline
     55-110 & 126-237 & 128-300 & 195-300 & - & - & - \\
     \hline
     60-105 & 138-228 & 139-300 & 213-300 & - & - & - \\
     \hline
     \multicolumn{7}{|c|}{\texttt{PARSEC-II}}\\
     \hline
     45-120 & 107-229 & 112-239 & 92-221 & 111-294 & 133-300 & - \\
     \hline
     55-110 & 117-150 & 130-227 & 109-202 & 138-270 & 166-300 & - \\
                & 153-211 &  &  &  &  &  \\
     \hline
     60-105 & 125-145 & 140-221 & 118-193 & 151-258 & 182-300 & - \\
                & 158-203 &  &  &  &  & \\
     \hline
     \multicolumn{7}{|c|}{\texttt{FRANEC}}\\
     \hline
     45-120 & 111-262 & 113-272 & 136-300 & 183-300 & 220-300 & - \\
     \hline
     55-110 & 131-242 & 134-251 & 173-300 & 233-300 & 282-300 & - \\
     \hline
     60-105 & 141-232 & 145-240 & 192-300 & 259-300 & - & - \\
     \hline
     \label{tab:mzams_intervals}
 \end{tabular}
\end{table*}

\begin{figure}
	\includegraphics[width=\columnwidth]{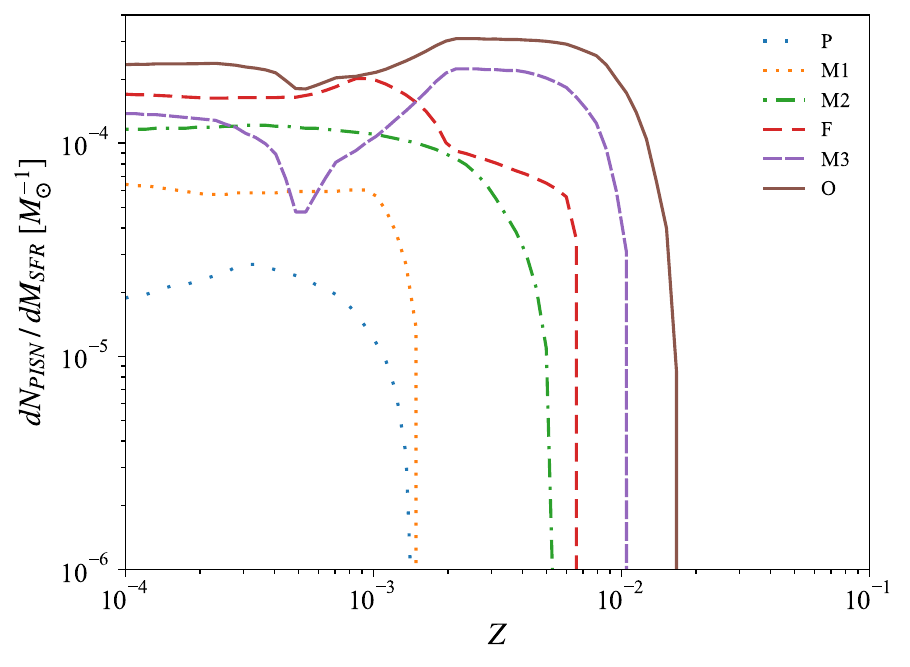}
    \caption{Number of PISN produced per unit star forming mass as a function of metallicity, for different combinations of stellar evolution code, CO core mass criterion and IMF upper limit (see Table \ref{tab:stellar_combos}). This figure clearly shows the maximum metallicity at which a star can explode as PISN, $Z_{\rm max}$, according to each variation.}
    \label{fig:pisnpermass}
\end{figure}

In Figure \ref{fig:pisnpermass}, we report the $dN_{\rm PISN}/dM_{\rm SFR}(Z)$ obtained using each stellar code, combined with different variations on $M_{\rm CO}$ criterion and $M_{\rm up}$. To avoid redundancy, among the eighteen cases that we compute, we decide to select six representative combinations, described in Table \ref{tab:stellar_combos}. \textit{F} stands for our fiducial variation. \textit{P} represents our pessimistic case, producing the smallest amount of PISNe, while \textit{O} is our optimistic case. All variations in the middle are marked with an $M$. One can clearly see the $Z_{\rm max}$ resulting from each combination, as the metallicity at which $dN_{\rm PISN}/dM_{\rm SFR}(Z)$ vanishes. The exact values are reported in Table \ref{tab:stellar_combos}. The effects of combining different stellar codes, $M_{\rm CO}$ range and $M_{\rm up}$, can be boiled down to $Z_{\rm max}$, and secondly to the height of the $dN_{\rm PISN}/dM_{\rm SFR}$ curve. For this reason, when presenting our results in Section \ref{sec:results}, every stellar variation will be represented by the corresponding $Z_{\rm max}$.

We note that some different combinations would lead to results superimposing to the ones we show, due to the degeneracy of the PISN rate on stellar prescriptions, but falling in any case between our \textit{P} and \textit{O} variations. 

\begin{table}
\centering
\captionof{table}{Considered variations on stellar evolution code, CO core mass criterion and IMF upper limit, as reference for Figure \ref{fig:pisnpermass}. The maximum metallicity to have PISN according to each variation, $Z_{\rm max}$, is also shown.}
\begin{tabular}{|c|c|c|c|c|}
 \hline
 name & stellar code & $M_{\rm CO}/M_{\rm \odot}$ & $M_{\rm up}/M_{\rm \odot}$ & $Z_{\rm max}$ \\
 \hline
 P & \texttt{FRANEC} & 60-105 & 150 & 1.5$\times 10^{-3}$ \\
 \hline
 M1 & \texttt{PARSEC-I} & 55-110 & 150 & 1.5$\times 10^{-3}$ \\
 \hline
 M2 & \texttt{FRANEC} & 45-120 & 150 & 5.5$\times 10^{-3}$ \\
 \hline
 F & \texttt{PARSEC-I} & 55-110 & 300 & 6.6$\times 10^{-3}$ \\
 \hline
 M3 & \texttt{PARSEC-II} & 45-120 & 150 & 1.0$\times 10^{-2}$ \\
 \hline
 O & \texttt{PARSEC-II} & 45-120 & 300 & 1.7$\times 10^{-2}$ \\
 \hline
 \label{tab:stellar_combos}
\end{tabular}
\end{table}

\subsection{Rate computation}\label{sec:rate_comp}

The galactic and stellar parts of our model provide us with the $Z$-dependent SFRD, $d^3M_{\rm SFR}/dtdVd\log Z\:(Z,z)$, and the number of PISNe produced per unit star forming mass, $dN_{\rm PISN}/dM_{\rm SFR}\:(Z)$, respectively. In order to finally compute the PISN event rate as a function of redshift, $d^2N_{\rm PISN}/dtdV\:(z)$, we need to convolve these two quantities by integrating over $Z$, as shown in Equation \ref{eq:rate_z}.

In Section \ref{sec:gal_prop}, we study the dependence of the PISN rate on galaxy stellar mass and metallicity, $M_{\rm \star}$ and $Z$. In order to do that, we need to start from the SFRD defined also per unit $M_{\rm \star}$, which we obtain by simply avoiding integrating over this quantity in Equation \ref{eq:CSFRD}. We then obtain the dependence on $M_{\rm \star}$ and $Z$ by integrating over $z$, after converting the unit volume into unit redshift via the comoving volume element per unit redshift and steradian, $dV/dz$:
\begin{equation}\label{eq:dV_to_dz}
\begin{split}
&\frac{d^4M_{\rm SFR}}{dzdtd\log Zd\log M_{\rm \star}}(Z,M_{\rm \star},z)=\\
&=\frac{d^4M_{\rm SFR}}{dtdVd\log Zd\log M_{\rm \star}}(Z,M_{\rm \star},z)\times 4\pi \frac{dV}{dz}(z),
\end{split}
\end{equation}
\begin{equation}\label{eq:rate_M_Z}
\begin{split}
&\frac{d^3N_{\rm PISN}}{dtd\log Zd\log M_{\rm \star}}(Z,M_{\rm \star})=\\
&=\int dz\:\frac{d^4M_{\rm SFR}}{dzdtd\log Zd\log M_{\rm \star}}(Z,M_{\rm \star},z)\times\frac{dN_{\rm PISN}}{dM_{\rm SFR}}(Z).
\end{split}
\end{equation}
By further integrating, one can get the PISN rate as a function of $M_{\rm \star}$ or $Z$. The possibility to handle the masses of single galaxies represents one of the main advantages of SFRD semi-empirical determinations like the one employed in this work. In particular, here it allows us to identify the masses of the favourable PISN host galaxies, besides their metallicity, as we show in Section \ref{sec:gal_prop}.

\section{Results}\label{sec:results}

In the following Section, we present the results obtained in this work. First, we show how our variations on stellar evolution code, CO core mass criterion and IMF upper limit, affect the PISN rate (Section \ref{sec:results_stellar}). Then we focus on the interplay between $Z_{\rm max}$, resulting from each stellar variation, and $\sigma_{\rm Z}$ in determining the PISN rate, and the effect of changing $\alpha_{\rm SMF}$ (Section \ref{sec:gal_var}). The aim is to study to what extent the uncertainties in both stellar and galactic models affect the PISN rate, and obtain a range of results spanning from a pessimistic to an optimistic case. 

Differently from PISNe, several hundreds of CCSNe have been observed so far. Thus it can be useful to compare the rates of these transients. We compute the ratio between PISN and CCSN rate, and show the results in Section \ref{sec:PI_CC_ratio}, for all of our variations.

Finally, in Section \ref{sec:gal_prop} we study the dependence of the PISN rate on galaxy $M_{\rm \star}$ and $Z$, in order to understand which galactic environments are favourable for PISN production, and how much galaxies with given properties contribute to the PISN rate. 
 
\subsection{Stellar variations} \label{sec:results_stellar}

\begin{figure*}
	\includegraphics[width=\columnwidth]{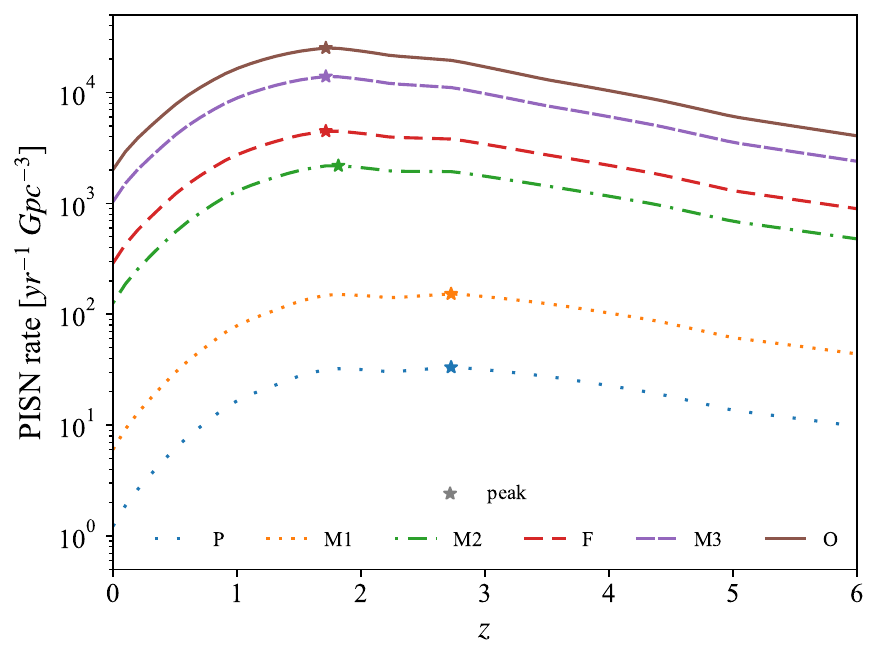}
    \caption{PISN rate as a function of $z$ for our stellar variations on stellar evolution code, CO core mass criterion and IMF upper limit. See Table \ref{tab:stellar_combos} for a description of each variation. Stars indicate the peak of the PISN rate.}
    \label{fig:var_stellar}
\end{figure*}

Figure \ref{fig:var_stellar} shows the PISN rate density as a function of redshift, computed with Equation \ref{eq:rate_z} for the stellar variations described in Section \ref{sec:st_ev_model}. The parameters of the galactic model are fixed to $\sigma_{\rm Z}=0.35$ and $\alpha_{\rm SMF}=-1.45$, which can be considered our fiducial values. The rate closely resembles the trend of the SFRD (Figure \ref{fig:CSFRD}), peaking at around $z=2$ and smoothly declining towards higher redshifts. 

We can see that our whole range of stellar variations, from pessimistic (\textit{P}) to optimistic (\textit{O}), leads to roughly three orders of magnitude in the PISN rate, with values at $z=0$ from $\sim 10^0$ to $10^3$ $Gpc^{-3}$ $yr^{-1}$, and values at peak from $\sim 3\times 10^1$ to $2\times 10^4$ $Gpc^{-3}$ $yr^{-1}$. As can be easily understood, a larger $M_{\rm CO}$ interval produces larger $M_{\rm ZAMS}$ ranges for the progenitors, i.e. more stars going into PISN. Analogously, extending the IMF from $M_{\rm up}=150$ to $300\:M_{\rm \odot}$ adds the contribution from more massive stars. This translates into higher $dN_{\rm PISN}/dM_{\rm SFR}$, with increased $Z_{\rm max}$, allowing for the production of more PISNe, at higher metallicities (Figure \ref{fig:pisnpermass}). As a result, one gets a higher PISN rate. E.g. if we fix \texttt{FRANEC} as stellar evolution code and $M_{\rm up}=150\:M_{\rm \odot}$, changing the $M_{\rm CO}$ criterion from $[60-105]$ to $[45-120]$ (i.e. variation $P$ to $M2$) increases the rate by about two orders of magnitude. Selecting instead \texttt{PARSEC-I}, $M_{\rm CO}\in[55-110]$, and bringing $M_{\rm up}$ from 150 to $300\:M_{\rm \odot}$ ($M1$ to $F$), leads to between one and two orders of magnitude increase in the rate.

The effects described above are weaker going towards higher PISN rates. E.g. if we consider \texttt{PARSEC-II} with $M_{\rm CO}\in[45-120]$, bringing $M_{\rm up}$ from 150 to 300 $M_{\rm \odot}$ ($M3$ to $O$) only leads to a factor $\sim 2$ difference in the rate. This can be understood by looking at Figure \ref{fig:rate_Z_z}, where we present the PISN rate distribution in the $z$-$Z$ plane for variations $P$, $F$ and $O$ (lower, middle and upper panels respectively). We decide to only select these variations for simplicity, being representative of the whole set. In particular, the central column shows results for $\sigma_{\rm Z}=0.35$, considered here (other variations on $\sigma_{\rm Z}$ are discussed in the next Section). These plots exhibit a sharp cut in the metallicity distribution, which is a direct effect of $Z_{\rm max}$, completely suppressing PISNe at higher metallicities. Variation $O$ displays $Z_{\rm max}\gtrsim 10^{-2}$, thus including the majority of the contribution from the SFRD peak, located at metallicities just above or below $10^{-2}$ depending on redshift (see the white lines in Figure \ref{fig:CSFRD_in_Z}). This can be best appreciated in Figure \ref{fig:rate_Z_z} by comparing the PISN rate distribution with the SFRD contour levels, indicated by violet dashed lines. Also variation $M3$ includes a fair portion of the SFRD peak contribution, with only a sligthly smaller $Z_{\rm max}$. Together with the fact that $dN_{\rm PISN}/dM_{\rm SFR}$ changes just by a factor less than 2, this causes the rate to increase only by a relatively small amount going from $M3$ to $O$. On the contrary, the PISN rate distribution for the more pessimistic variations ($P$ and $M1$) extends to metallicities far below the SFRD peak. As a consequence, moving to variations $M2$ and $F$ adds a significant contribution to the PISN rate, from metallicities closer to the SFRD peak. This effect is further enhanced by the stronger increase in $dN_{\rm PISN}/dM_{\rm SFR}$, especially from $P$ to $M2$, with respect to the more optimistic cases. 

As indicated by the stars in Figure \ref{fig:var_stellar}, also the peak of the PISN rate distribution in redshift appears to shift with stellar variations, an effect which is enhanced when combining with galactic variations (see the following Section). It is due to the fact that, on average, metallicity tends to decrease with redshift, according to our $Z$-evolution recipe. Variations with lower $Z_{\rm max}$ thus favour the contribution to the PISN rate coming from high redshift. As a result, the position of the peak tends to shift towards higher redshifts, always below $z=3$ considering only stellar variations. 

\subsection{Galactic variations} \label{sec:gal_var}

After showing how different stellar evolution prescriptions, and IMF upper limits, influence the PISN rate, we now focus on the variations on parameters $\sigma_{\rm Z}$ and $\alpha_{\rm SMF}$ of our galactic model (see Section \ref{sec:gal_model}). 

\begin{figure*}
	\includegraphics[scale=0.5]{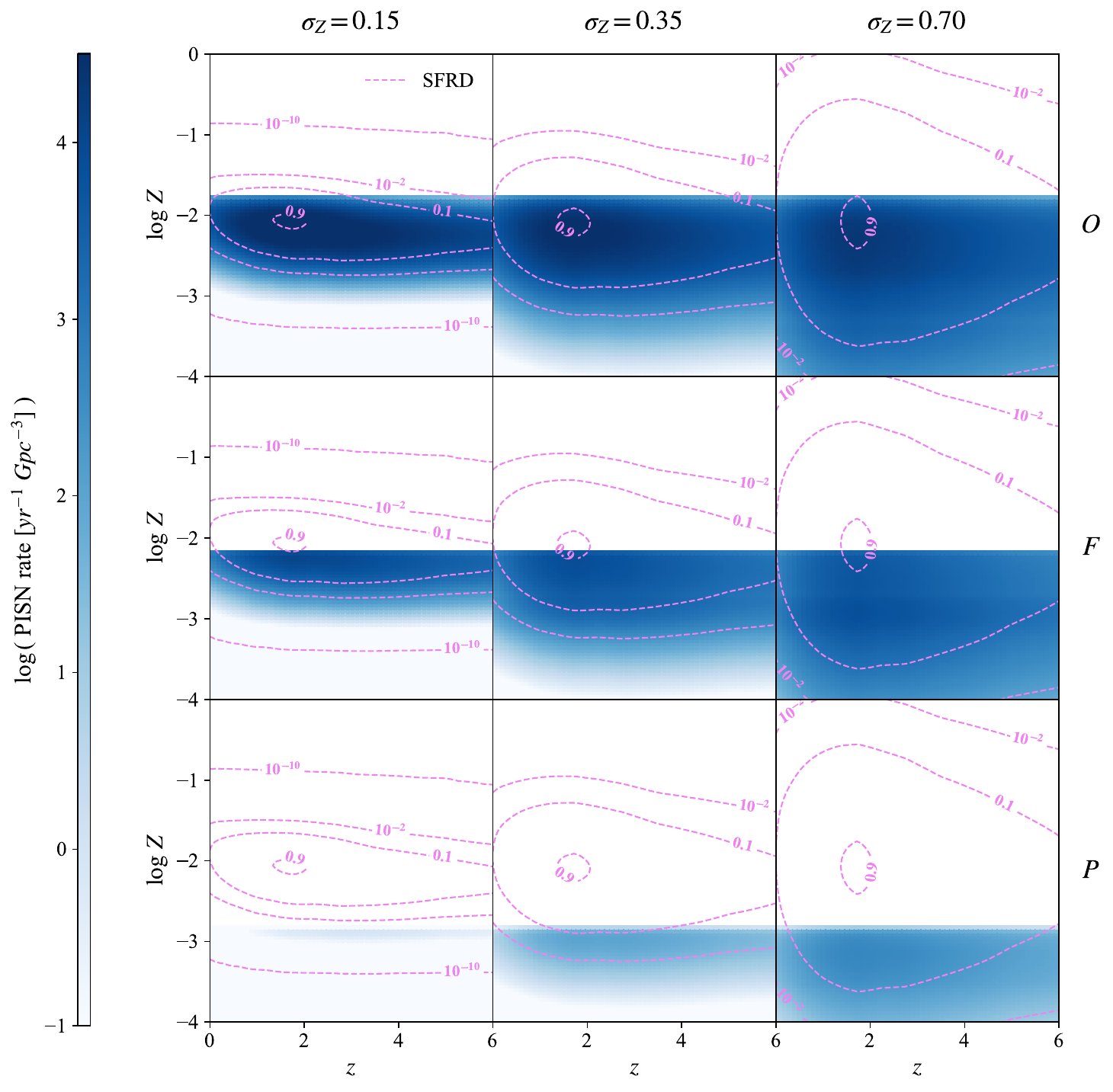}
    \caption{PISN rate as a function $z$ and $Z$ for variations $P$, $F$ and $O$ on stellar evolution code, $M_{\rm CO}$ criterion and $M_{\rm up}$ (lower, middle and upper panels respectively). See Table \ref{tab:stellar_combos} for a description of each variation, and the corresponding $Z_{\rm max}$. Different columns correspond to variations $\sigma_{\rm Z}=0.15$, 0.35, 0.70 (left to right). We also show the SFRD contour levels for comparison (violet dashed lines), corresponding to fractions of $10^{-10}$, $10^{-2}$, 0.1 and 0.9 times its peak value.}
    \label{fig:rate_Z_z}
\end{figure*}

\begin{figure*}
	\includegraphics[width=\columnwidth]{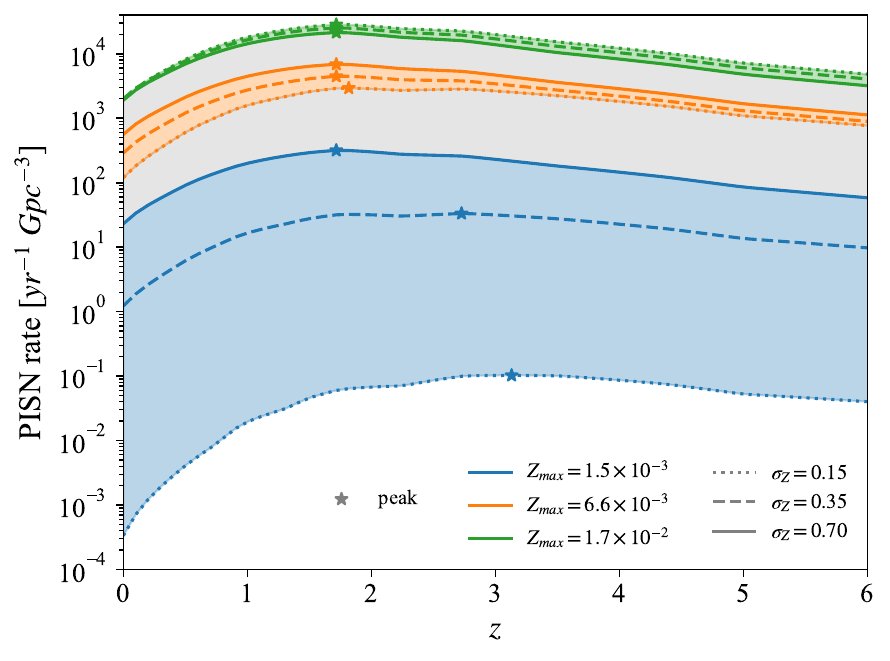}
    \caption{PISN rate as a function of $z$ for different $\sigma_{\rm Z}$, represented by different linestyles, for $Z_{\rm max}$ corresponding to variations $P$, $F$ and $O$ (blue, orange and green lines, see Table \ref{tab:stellar_combos}). Bands of the corresponding colors show the range of results produced by varying $\sigma_{\rm Z}$, for every fixed $Z_{\rm max}$. Stars indicate the peak of the PISN rate. We also plot a grey band to stress that, for clarity reasons, here we do not show all possible combinations of our model variations, but the ones we left out would fill in the gaps between, and be included within, the representative ones that we selected.}
    \label{fig:Zmax_sigmaZ}
\end{figure*}

As anticipated above, we find a strong interplay between $Z_{\rm max}$, resulting from the combination of stellar evolution code, $M_{\rm CO}$ criterion and $M_{\rm up}$, and $\sigma_{\rm Z}$. Among the stellar variations discussed in the previous Section, for clarity we decide to select only the most pessimistic and optimistic cases ($P$ and $O$), as well as the fiducial one ($F$). Figure \ref{fig:Zmax_sigmaZ} shows the PISN rates we obtain for the combination of each of these variations, described by the corresponding $Z_{\rm max}$ (see Table \ref{tab:stellar_combos}), with different $\sigma_{\rm Z}$. As we can see, this results in PISN rates spanning roughly seven orders of magnitude considering values at $z=0$, or five orders of magnitude considering values at $z=6$. Most of this range is due to variations with lowest $Z_{\rm max}=1.5\times 10^{-3}$. Indeed, as already discussed in the previous Section and shown in Figure \ref{fig:rate_Z_z}, such low $Z_{\rm max}$ removes the main contribution to the PISN rate, coming from metallicities close to the SFRD peak. Selecting a low $\sigma_{\rm Z}$ means considering a SFRD which does not extend to the lowest metallicities (Figure \ref{fig:CSFRD_in_Z}, left panel), and as a consequence it strongly suppresses the PISN rate contribution from the tail of the galaxy metallicity distribution, $Z<Z_{\rm max}$. This is the reason why variation $P$ is so dependent on $\sigma_{\rm Z}$. On the other hand, variation F yields a $Z_{\rm max}=6.6\times 10^{-3}$ which is closer to the SFRD peak, and even in the lowest $\sigma_{\rm Z}$ case it includes part of that contribution (middle panels in Figure \ref{fig:rate_Z_z}). Therefore the rate suffers less dramatically from changes in $\sigma_{\rm Z}$. Variation $O$, with $Z_{\rm max}=1.7\times 10^{-2}$ (top panels in Figure \ref{fig:rate_Z_z}), includes most of the contribution from the SFRD peak, making the dependence of the PISN rate on $\sigma_{\rm Z}$ even fainter. For all $Z_{\rm max}$, this dependence appears more enhanced at lower redshifts. This is due to the fact that, as shown in Figure \ref{fig:CSFRD_in_Z}, our SFRD experiences a substantial decrease below $z\sim 1$, resulting in a drop in the rate which is more and more significant going to lower $\sigma_{\rm Z}$. 

In variations $P$ and $F$, the PISN rate increases with $\sigma_{\rm Z}$. Indeed, in these cases $Z_{\rm max}$ lies below the peak of the SFRD, therefore a higher dispersion for the galaxy metallicity distribution increases the PISN rate below $Z_{\rm max}$. On the contrary, variation $O$ exhibits a rate increasing with decreasing $\sigma_{\rm Z}$ (namely, the dotted line is above the solid one). This is because in this case the peak of the SFRD is already included below $Z_{\rm max}$. A higher $\sigma_{\rm Z}$ thus spreads a fraction of the PISN rate distribution above $Z_{\rm max}$, where it gets lost (Figure \ref{fig:rate_Z_z}). 

All in all, the PISN rate dependence on $\sigma_{\rm Z}$ varies dramatically with $Z_{\rm max}$, increasing significantly going to lower values of this parameter (see the color bands in Figure \ref{fig:Zmax_sigmaZ}, which get larger for lower $Z_{\rm max}$). This shows the strong interplay between these parameters, which turns out to be crucial in determining the PISN rate. It is to be noted that, if we applied our galactic variations to the other stellar variations ($M1$, $M2$ and $M3$), we would obtain ranges of results partially superimposing to the ones we show for variations $P$, $F$ and $O$, as expressed by the grey band in Figure \ref{fig:Zmax_sigmaZ}. This reveals the degeneracy of the PISN rate on stellar and galactic prescriptions.

One can also notice a change in the peak position, from $z\sim 2$ (reproducing the SFRD peak) for $\sigma_{\rm Z}=0.70$, to $z\gtrsim 3$ for $\sigma_{\rm Z}=0.15$, in the lowest $Z_{\rm max}=1.5\times 10^{-3}$ variation. As explained in the previous section, this is an effect of the average metallicity decrease with redshift, prescribed by our Z evolution recipe. Here this effect is enhanced by the $\sigma_{\rm Z}=0.15$ variation, which further favours the PISN rate contribution from higher $z$.

\begin{figure*}
	\includegraphics[width=\columnwidth]{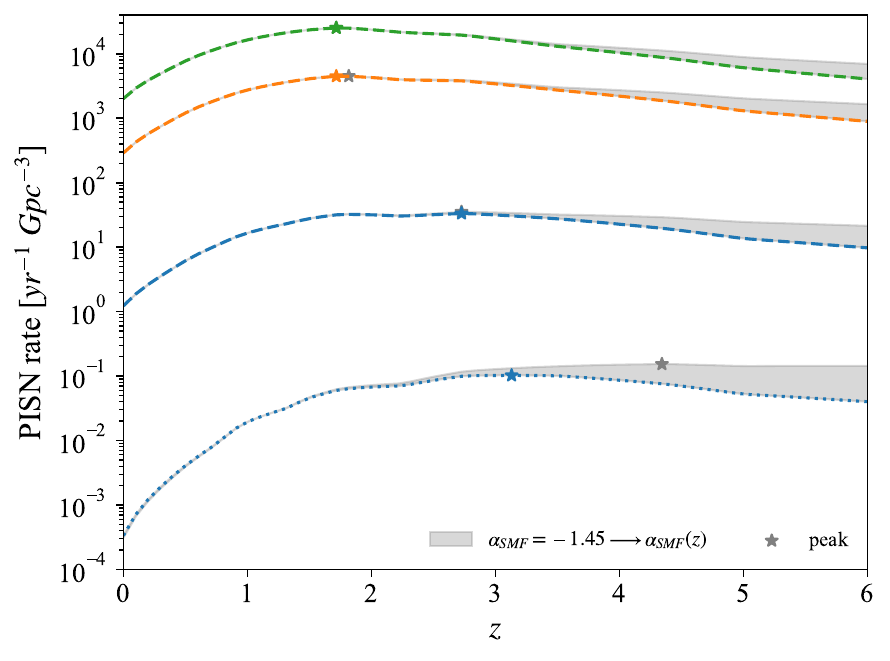}
    \caption{PISN rate as a function of $z$ for $\alpha_{\rm SMF}=-1.45$ and $\alpha_{\rm SMF}=\alpha_{\rm SMF}(z)$ variations. We show the difference between the two variations with a grey band. Lines are as in Figure \ref{fig:Zmax_sigmaZ} (note that we only reported some of those variations, for clarity). Stars indicate the peak of the PISN rate for each variation.}
    \label{fig:alpha_SMF}
\end{figure*}

Figure \ref{fig:alpha_SMF} shows the effect of varying $\alpha_{\rm SMF}$ prescription. The grey band represents the difference between the constant $\alpha_{\rm SMF}$ case, $\alpha_{\rm SMF}=-1.45$, and that with $\alpha_{\rm SMF}$ defined as a function of redshift, $\alpha_{\rm SMF}=\alpha_{\rm SMF}(z)$. We consider only some of the variations reported in Figure \ref{fig:Zmax_sigmaZ}, for clarity purposes, since the trend is analogous. We can see that prescribing a redshift dependence for $\alpha_{\rm SMF}$ has an appreciable effect only at high redshifts, $z>3-4$, where it increases the rate by a factor in any case smaller than one order of magnitude. Indeed, as shown in Figure \ref{fig:gsmf}, this variation increases the number of galaxies at low masses, an effect which is stronger going to higher redshift. Moreover, the FMR prescribes that low-mass galaxies are also metal poor. Since this relation produces an average metallicity decreasing with redshift, these low-mass galaxies end up increasing the PISN rate at high redshift. This can also be appreciated by looking at Figure \ref{fig:CSFRD_in_Z}, where one can see that the $Z$ corresponding to the peak of the SFRD rate experiences a steeper decrease with redshift for variation $\alpha_{\rm SMF}(z)$, with respect to constant $\alpha_{\rm SMF}$ (an effect already shown in \citealt{Chruslinska_2019}). For the same reason, in variation $P$ the peak of the $\alpha_{\rm SMF}(z)$ rate is shifted to between $z=4$ and 5. Indeed, the effects of the $Z_{\rm max}-\sigma_{\rm Z}$ dependence are exacerbated, and in particular the PISN rate peak is brought to even higher redshifts than the constant-$\alpha_{\rm SMF}$ case.

We warn that the huge range we obtain for the PISN rate, by accounting for both stellar and galactic variations, is actually strongly dependent on the adopted $Z$-evolution prescriptions. We show this point in Section \ref{sec:FMR_var}, where we explore an additional variation on FMR.

\begin{figure}
	\includegraphics[width=\columnwidth]{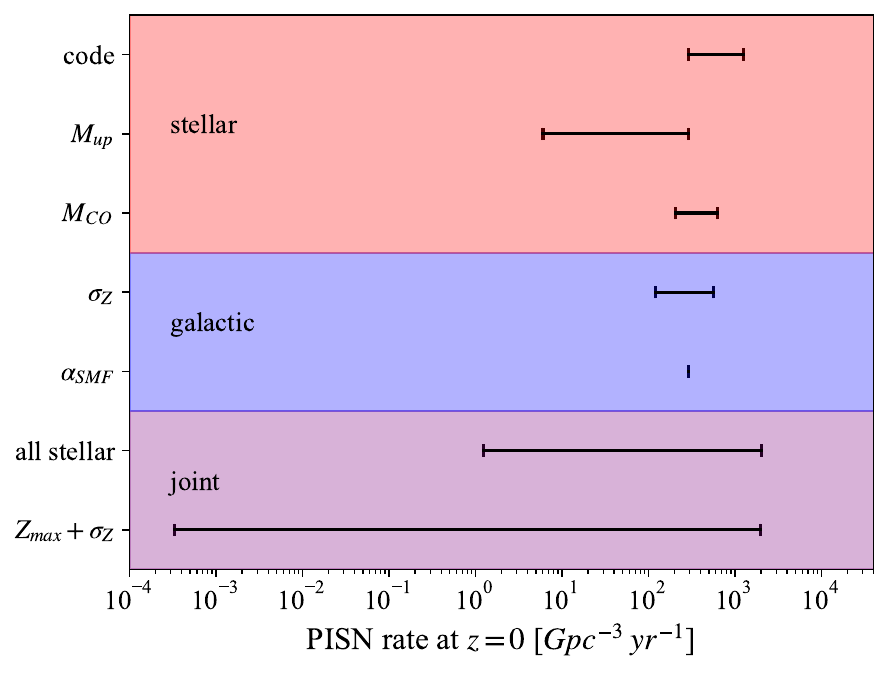}
    \caption{Uncertainties on the local PISN rate, computed at $z=0$, due to each considered variation on stellar and galactic prescriptions, indicated with a red and blue band respectively (see the text). Global uncertainties due to all stellar variations combined are also shown, as well as those due to all possible combinations of $Z_{\rm max}$ and $\sigma_{\rm Z}$ (purple band). Indeed, as highlighted in the text, the interplay between these two parameters plays a crucial role in determining the PISN rate. This figure was inspired by Figure 5 in \citealt{Farmer_2019}.}
    \label{fig:uncertainty_intervals}
\end{figure}

Finally, Figure \ref{fig:uncertainty_intervals} shows the uncertainty intervals on the local PISN rate, computed at $z=0$, due to each individual stellar and galactic variation (red and blue bands respectively). Uncertainties due to all stellar variations, as well as all possible $Z_{\rm max} - \sigma_{\rm Z}$ combinations, are also shown (purple band). Every bar has been computed by fixing all other prescriptions to the fiducial case, and varying only the quantity of interest over the whole range. Notice how the $\alpha_{\rm SMF}$ variation does not lead to significant uncertainties on the PISN rate at $z=0$ since, as explained in this Section, it produces appreciable effects only at $z\sim 3-4$. This Figure clearly shows how crucial the interplay between $Z_{\rm max}$ and $\sigma_{\rm Z}$ is in determining the PISN rate, extending the possible values downwards by several orders of magnitude, with respect to considering stellar and galactic variations separately.

\subsection{PISN over CCSN ratio}\label{sec:PI_CC_ratio}

In this Section, we compute the ratio between the PISN and CCSN rate as a function of redshift. We follow the same procedure outlined in Section \ref{sec:model}, fixing $\alpha_{\rm SMF}=-1.45$. We modify the mass range for IMF integration in Equation \ref{eq:stellar_part} to [8-50] $M_{\rm \odot}$. The upper ZAMS-mass limit of CCSN progenitors is highly uncertain, possibly ranging from $\sim 20-25\:M_{\odot}$ to $125\:M_{\odot}$ (\citealt{Heger_2002,Heger_2003,Dahlen_2004,Cappellaro_2005,Botticella_2007,Smartt_2009,Botticella_2012,Dahlen_2012,Mattila_2012,Melinder_2012,Strolger_2015,Petrushevska_2016,Ziegler_2022}). 
For this reason, we select an intermediate, fiducial value of $50\:M_{\odot}$. We note that varying this value along the credible range does not change our results significantly, due to our bottom-heavy IMF. It is also to be stressed that, while for PISNe we obtain a dependence of the $M_{\rm ZAMS}$ range on metallicity, here we keep it constant, since the CCSN mass range is not expected to exhibit such a crucial dependence (namely CCSNe are expected to take place at all metallicities).

The results are shown in Table \ref{tab:PI_CC}, for the same variations as Figure \ref{fig:Zmax_sigmaZ}. Analogously to the PISN rate, the PI/CC ratio spans from $\sim$ five to less than seven orders of magnitude, depending on redshift. It ranges from $2.5\times 10^{-9}$ to $1.5\times 10^{-2}$ at $z=0$, and from $2.3\times 10^{-7}$ to $2.8\times 10^{-2}$ at $z=6$. Considering the redshift at which the PISN rate peaks for each variation, $z_{\rm peak}^{\rm PI}$, we get a PI/CC ratio ranging from $1.4\times 10^{-7}$ to $2.2\times 10^{-2}$. The PISN rate dependence on $Z_{\rm max} - \sigma_{\rm Z}$ combinations is closely reproduced. This is easily understood, since the CCSN rate consists simply in the multiplication of the SFRD by the constant factor $dN_{\rm CCSN}/dM_{\rm SFR}$. Therefore, all the features of the PISN rate, which are ultimately due to its metallicity dependence, are again found in the PI/CC ratio. Values at $z=6$ are always higher than at $z=0$ and even $z_{\rm peak}^{\rm PI}$. This is because PISNe occur in environments with $Z<Z_{\rm max}$, which are found preferentially at high redshift, according to our $Z$-evolution recipe. Since CCSNe are independent from metallicity, this favours PISNe over CCSNe, and causes their ratio to increase with redshift. The PISN rate drop at low redshifts, which is again due to the metallicity dependence and is thus absent in the CCSN rate, further accentuates this feature.

\begin{table}
\centering
\captionof{table}{PISN over CCSN rate ratio for our set of $Z_{\rm max}-\sigma_{\rm Z}$ combinations (same as Figure \ref{fig:Zmax_sigmaZ}). We report values at $z=0$ and $z=6$, as well as the redshift at which the PISN rate peaks ($z_{\rm peak}^{\rm PI}$), in order to show the whole range across redshift.}
\begin{tabular}{|c|c|c|c|}
 \hline
 $\sigma_{\rm Z}$ & PI/CC ($z=0$) & PI/CC ($z=z_{\rm peak}^{\rm PI}$) & PI/CC ($z=6$)\\
 \hline
 \multicolumn{4}{|c|}{variation $P$ ($Z_{\rm max}=1.5\times 10^{-3}$)}\\
 \hline
 0.15 & $2.5\times 10^{-9}$ & $1.4\times 10^{-7}$ & $2.3\times 10^{-7}$ \\
 \hline
 0.35 & $9.2\times 10^{-6}$ & $3.5\times 10^{-5}$ & $5.5\times 10^{-5}$ \\
 \hline
 0.70 & $1.7\times 10^{-4}$ & $2.4\times 10^{-4}$ & $3.3\times 10^{-4}$ \\
 \hline
 \multicolumn{4}{|c|}{variation $F$ ($Z_{\rm max}=6.6\times 10^{-3}$)}\\
 \hline
 0.15 & $9.2\times 10^{-4}$ & $2.3\times 10^{-3}$ & $4.5\times 10^{-3}$ \\
 \hline
 0.35 & $2.2\times 10^{-3}$ & $3.5\times 10^{-3}$ & $5.2\times 10^{-3}$ \\
 \hline
 0.70 & $4.3\times 10^{-3}$ & $5.4\times 10^{-3}$ & $6.6\times 10^{-3}$ \\
 \hline
 \multicolumn{4}{|c|}{variation O ($Z_{\rm max}=1.7\times 10^{-2}$)}\\
 \hline
 0.15 & $1.5\times 10^{-2}$ & $2.2\times 10^{-2}$ & $2.8\times 10^{-2}$ \\
 \hline
 0.35 & $1.5\times 10^{-2}$ & $2.0\times 10^{-2}$ & $2.4\times 10^{-2}$ \\
 \hline
 0.70 & $1.5\times 10^{-2}$ & $1.7\times 10^{-2}$ & $1.9\times 10^{-2}$ \\
 \hline
 \label{tab:PI_CC}
\end{tabular}
\end{table}

\subsection{Host galaxy properties}\label{sec:gal_prop}

In this Section we explore the dependencies of the PISN rate on galaxy mass and metallicity, and the interplay between them. Among our set of variations, we select three cases which best serve this purpose. The rates have been computed with Equations \ref{eq:dV_to_dz} and \ref{eq:rate_M_Z}, further integrating over $M_{\rm \star}$ or $Z$. We stress that, due to the integration over $z$ in Equation \ref{eq:rate_M_Z}, these rates represent the contribution coming from all redshifts up to $z=6$.

\begin{figure*}
        \centering
	\includegraphics[scale=0.55]{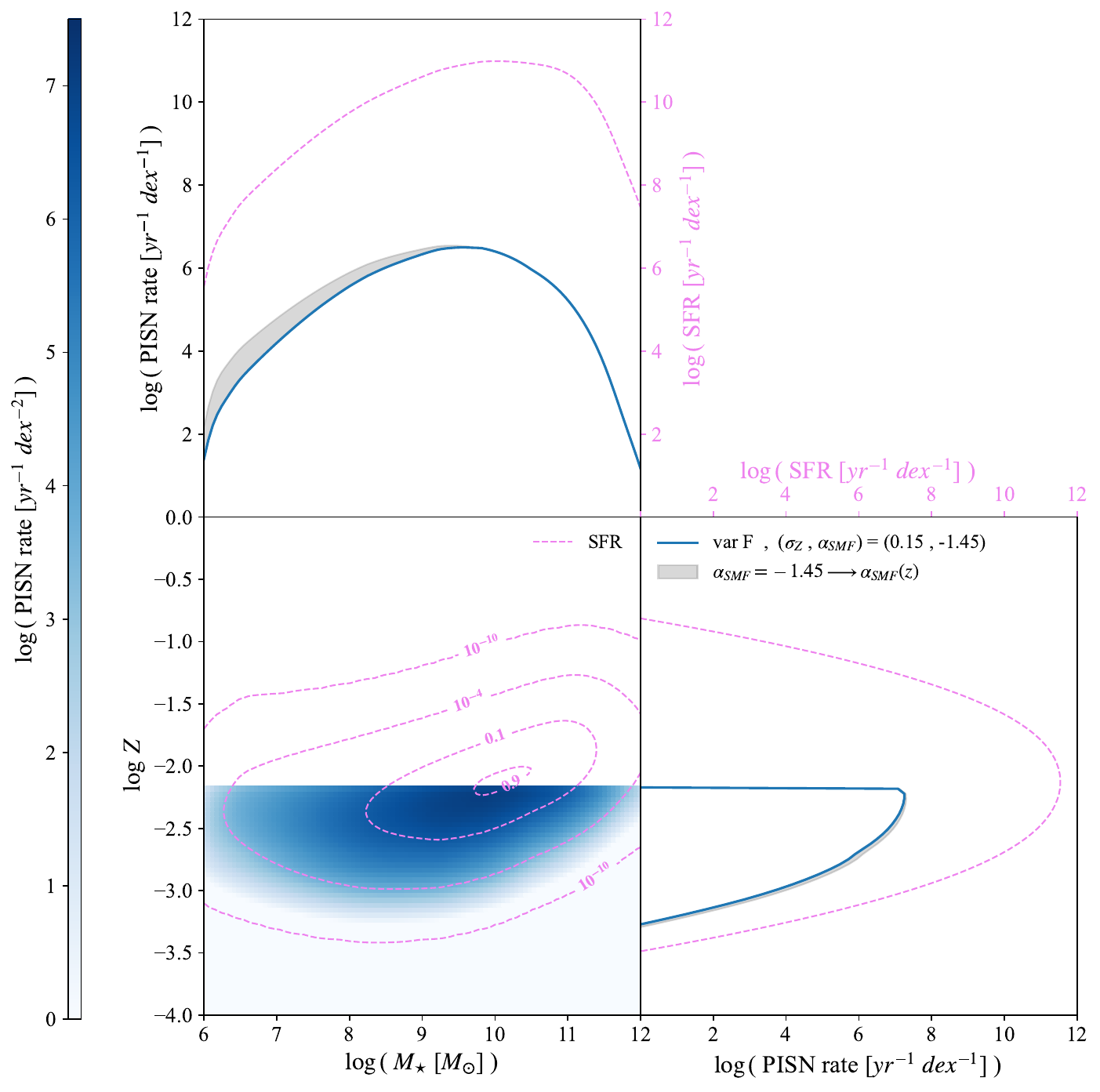}
    \caption{PISN rate as a function of $M_{\rm \star}$, $Z$ and both, for our fiducial stellar variation (\texttt{PARSEC-I}, $M_{\rm CO}\in[55-110]\:M_{\rm \odot}$, $M_{\rm IMF}^{\rm up}=300$ $M_{\rm \odot}$), with $\sigma_{\rm Z}=0.15$ and $\alpha_{\rm SMF}=-1.45$. Grey areas indicate the transition from the constant $\alpha_{\rm SMF}$ to the $\alpha_{\rm SMF}(z)$ case. For comparison, we also show the SFR contour levels as a function of $M_{\rm \star}$ and $Z$ for the corresponding variations, with fixed $\alpha_{\rm SMF}=-1.45$, as well as the individual dependencies of the SFR on $M_{\rm \star}$ and $Z$ (violet dashed lines). In particular, we show the contours at $10^{-10}$, $10^{-4}$, 0.1 and 0.9 the SFR peak. As stressed in the text, these rates represent the contribution coming from all redshifts up to $z=6$ (see Equation \ref{eq:rate_M_Z}).}
    \label{fig:host_gal_fiducial}
\end{figure*}

In Figure \ref{fig:host_gal_fiducial} we present a corner plot showing the individual and joint dependencies of the PISN rate on $M_{\rm \star}$ and $Z$, for our fiducial stellar variation $F$, with galactic parameter $\sigma_{\rm Z}=0.15$. The difference between the two GSMF low-mass end slope cases, $\alpha_{\rm SMF}=-1.45$ and $\alpha_{\rm SMF}=\alpha_{\rm SMF}(z)$, is indicated by a grey band. We find it informative to also show the SFR for the corresponding variations, with $\alpha_{\rm SMF}=-1.45$, telling us about the total star forming mass available in the first place. In particular, we show both the contour levels for the SFR as a function of $M_{\rm \star}$ and $Z$, as well as the individual dependencies of the SFR on each of these quantities (violet dashed lines). We find again the sharp cut in the metallicity distribution due to $Z_{\rm max}$, already discussed for Figure \ref{fig:rate_Z_z}. In this case $Z_{\rm max}=6.6\times 10^{-3}$, which lies below the peak of the SFRD, as indicated by the violet dashed contours in the bottom left panel (see also Figure \ref{fig:CSFRD_in_Z}). As a consequence, the main contribution to the PISN rate, coming from metallicities where the SFRD peaks, gets partially cut out. As discussed in Sections \ref{sec:results_stellar} and \ref{sec:gal_var}, this is the reason why the PISN rate turns out to be lower with respect to other variations with higher $Z_{\rm max}$. 

As one can see from the right panel of Figure \ref{fig:host_gal_fiducial}, the PISN rate peaks at metallicities just below $Z_{\rm max}$, as a result of combining $dN_{\rm PISN}/dM_{\rm SFR}$ (variation $F$ in Figure \ref{fig:pisnpermass}) with the SFRD. In other words, PISN production is favoured in low-$Z$ environments, but stars formed at $Z$ closer to the SFRD peak are more abundant, and therefore provide a larger contribution to the PISN rate.

In the top panel of Figure \ref{fig:host_gal_fiducial} we show the PISN rate dependence on $M_{\rm \star}$. This variation favours host galaxies with stellar masses between $10^9$ and $10^{10}$ $M_{\rm \odot}$, from which comes the main contribution to the PISN rate. These masses are somewhat lower than those at which the SFR peaks, around $\sim 10^{10}\:M_{\rm \odot}$, as indicated by the violet dashed line in the top panel. Indeed, partially cutting the peak of the SFRD distribution over metallicity, also stops the rise of the PISN rate with $M_{\rm \star}$, as can be clearly seen in the bottom left panel by comparing the PISN rate distribution with the SFR violet dashed contours. This is due to the correlation between $M_{\rm \star}$ and $Z$ prescribed by our FMR. 

\begin{figure*}
        \centering
	\includegraphics[scale=0.55]{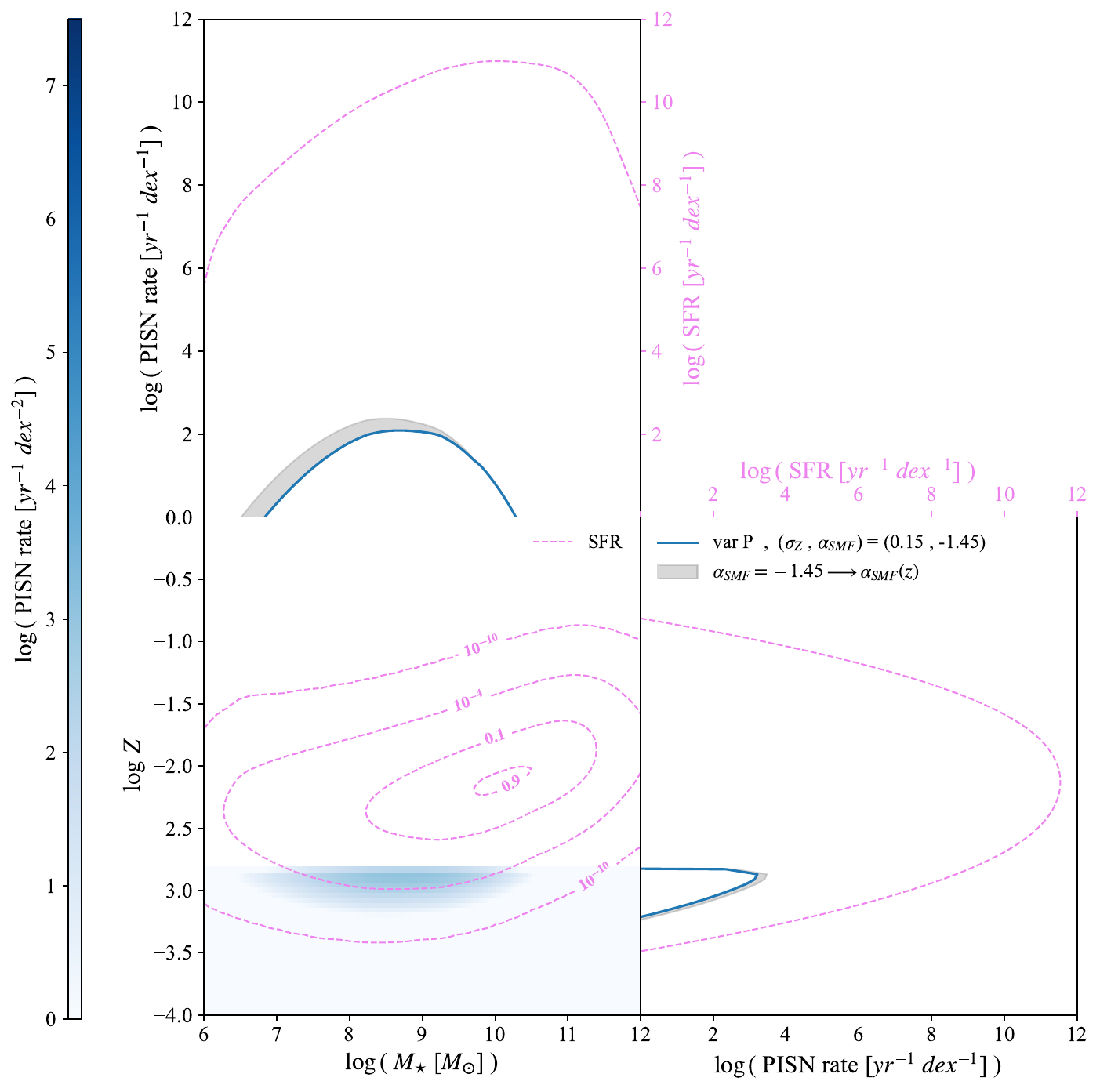}
    \caption{As in Figure \ref{fig:host_gal_fiducial}, for our pessimistic variation (\texttt{FRANEC}, $M_{\rm CO}\in[60-105]\:M_{\rm \odot}$, $M_{\rm IMF}^{\rm up}=150$ $M_{\rm \odot}$), with $\sigma_{\rm Z}=0.15$. Notice the different axis ranges for the PISN rate, due to the low values.}
    \label{fig:host_gal_pessimistic}
\end{figure*}

Figure \ref{fig:host_gal_pessimistic} shows the results obtained for variation $P$, with $\sigma_{\rm Z}=0.15$. This is the most pessimistic case considered in this work. Variation $P$ features the lowest $Z_{\rm max}=1.5\times 10^{-3}$, leading to an even more dramatic cut in the metallicity distribution, with respect to the previous case shown. As a consequence, the rate ends up being more than four orders of magnitude lower than in the previous case. Moreover, the rate peak shifts to a lower metallicity, just above $10^{-3}$, reflecting the trend of the $dN_{\rm PISN}/dM_{\rm SFR}$ over metallicity (variation $P$ in Figure \ref{fig:pisnpermass}). The mass distribution gets shifted towards lower values, favouring galaxies with $M_{\rm \star}$ between $10^8$ and $10^9\:M_{\rm \odot}$. This is again due to the correlation between $M_{\rm \star}$ and $Z$, given the lower $Z_{\rm max}$ cut. Since $Z_{\rm max}$ is far below the SFRD peak, a major contribution to the PISN rate is taken out of the game. This is the reason why in this variation the rate experiences such a dramatic decrease, as discussed in Sections \ref{sec:results_stellar} and \ref{sec:gal_var}.\\

\begin{figure*}
        \centering
	\includegraphics[scale=0.55]{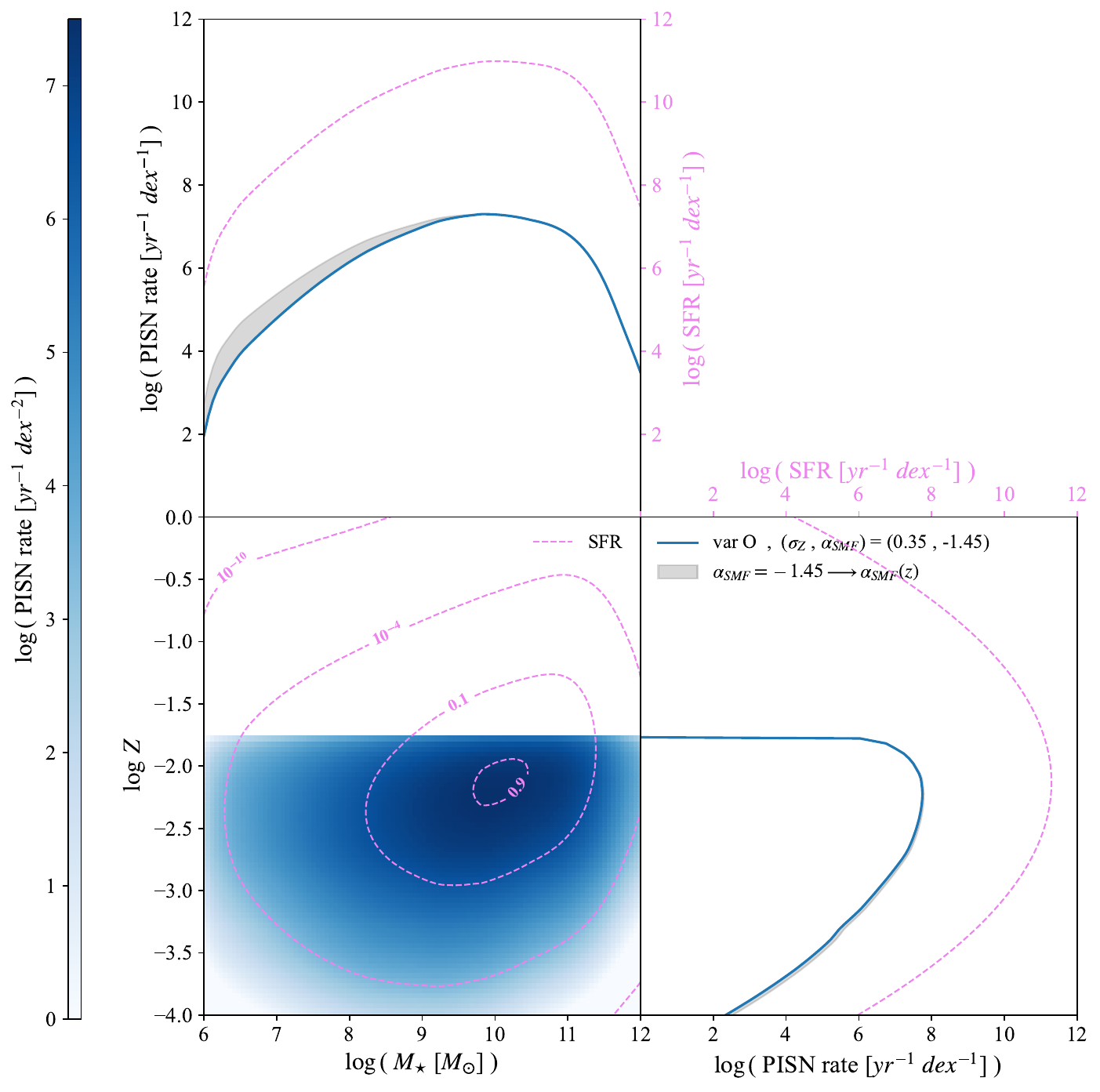}
    \caption{As in Figure \ref{fig:host_gal_fiducial}, for our optimistic variation (\texttt{PARSEC-II}, $M_{\rm CO}\in[45-120]\:M_{\rm \odot}$, $M_{\rm IMF}^{\rm up}=300$ $M_{\rm \odot}$), with $\sigma_{\rm Z}=0.35$.}
    \label{fig:host_gal_optimistic}
\end{figure*}

Finally, in Figure \ref{fig:host_gal_optimistic} we present the host galaxy properties obtained for variation $O$, combined with $\sigma_{\rm Z}=0.35$. As already discussed, here $Z_{\rm max}$ is greater than the SFRD peak, thus the rate is increased with respect to the fiducial case, by a factor less than one order of magnitude. Moreover, the peaks in both the $Z$ and $M_{\rm \star}$ distributions resemble the SFR ones since, differently from the previous cases shown, here the main contribution to the SFR is almost-fully included in the PISN rate distribution. In particular, the peak lies at $Z$ between $10^{-2}$ and $10^{-2.5}$, and $M_{\rm \star}\lesssim 10^{10}\:M_{\rm \odot}$.\\

In all cases, prescribing a redshift dependence of $\alpha_{\rm SMF}$ increases the rate at $M_{\rm \star}\lesssim 10^{10}\:M_{\rm \odot}$, as can be seen by the grey band in the top panel. Indeed, the $\alpha_{\rm SMF}(z)$ variation produces more galaxies with low mass with respect to the fixed $\alpha_{\rm SMF}$ case (Figure \ref{fig:gsmf}). Because of the correlation between $M_{\rm \star}$ and $Z$, these galaxies will also be metal poor, explaining why the PISN rate slightly increases at metallicities below the SFRD peak (grey band in the lower right panel).

\section{Discussion}\label{sec:discussion}

\subsection{PISN rate and variations}
Our results indicate that using different stellar evolution prescriptions, and varying the relevant parameters of the galactic model, hugely affects the PISN rate. Different stellar evolution codes, $M_{\rm CO}$ criteria and IMF upper limits produce a $\sim$ three orders of magnitude range from the most pessimistic to optimistic case (variations $P$ and $O$ in Figure \ref{fig:var_stellar}). These variations play a crucial role through the quantity $Z_{\rm max}$, and secondly through the height of the $dN_{\rm PISN}/dM_{\rm SFR}$ curve. As shown in Figure \ref{fig:Zmax_sigmaZ}, the interplay between $Z_{\rm max}$ and $\sigma_{\rm Z}$ broadens the PISN rate range to between $\sim$ five and seven orders of magnitude, throughout the whole redshift range considered. Indeed, $Z_{\rm max}$ completely cuts the SFRD distribution over metallicity, thus $\sigma_{\rm Z}$ strongly regulates the amount of SFRD which gets saved or lost (Figure \ref{fig:rate_Z_z}). This effect becomes huge for low $Z_{\rm max}$. E.g., in variation $P$, the peak of the PISN rate drops from $\sim 10^1$ to $10^{-1}\:Gpc^{-3}$ $yr^{-1}$, by decreasing $\sigma_{\rm Z}$ from 0.35 to 0.15. Moreover, the PISN rate peak is moved from $z\lesssim 2$ to $z\gtrsim 3$, for the most pessimistic variation. Finally, varying $\alpha_{\rm SMF}$ increases the PISN rate at $z>3-4$ by a factor less than one order of magnitude, and shifts its peak to even higher redshifts than the constant-$\alpha_{\rm SMF}$ case, up to $z\sim 4-5$ (Figure \ref{fig:alpha_SMF}).

These results show how the existing uncertainties in stellar and galaxy evolution hamper the determination of the cosmic PISN rate, with uncertainties dominated by the interplay between $Z_{\rm max}$ and $\sigma_{\rm Z}$. On the other hand, this offers the opportunity to constrain stellar and galactic models, thanks to possible PISN observations in the future. Also in the eventuality that PISNe are never discovered, their lack of observations would help pose limits on the models' parameters. 

According to our PISN rate distributions over galactic $M_{\rm \star}$ and $Z$, the stellar mass of the favourable PISN host galaxies ranges from $10^8-10^9\:M_{\rm \odot}$, to around $10^{10}\:M_{\rm \odot}$, going from pessimistic to optimistic variation. Favoured galaxy metallicities range from $\sim 10^{-3}$ to just below $10^{-2}$, following $Z_{\rm max}$. These results tell us about the galactic environments which are most favourable to host PISNe, and can be taken as indication for current and future surveys aimed at observing these transients. 

\subsection{Comparison with previous works}
It is useful to compare our results with previous works that, similarly to us, compute the PISN rate by combining a stellar evolution code with a cosmic star formation history and $Z$-evolution recipe (\citealt{Stevenson_2019,du_Buisson_2020,Briel_2022_1,Hendriks_2023,Tanikawa_2022,Tanikawa_2024}). We note that, in all these comparisons, stellar and galactic prescriptions were selected in order to be as compatible as possible. Moreover, these works also consider PISNe arising from binary stars, while we only consider single stars. As we will discuss in the next Section, we do not expect taking binaries into account to influence the PISN rate significantly. The $Z$-dependent SFRDs adopted in the following works exhibit $\sigma_{\rm Z}$ equal, or roughly compatible, to our $\sigma_{\rm Z}=0.35$ choice. Therefore, we compare with the results shown in Figure \ref{fig:var_stellar}, where we apply stellar variations keeping $\sigma_{\rm Z}=0.35$ fixed. 

Overall, these previous determinations tend to fall in the lower end of our Figure \ref{fig:var_stellar} range. This is partially due to the employed stellar evolution codes, producing different $dN_{\rm PISN}/dM_{\rm SFR}(Z)$ with respect to our work. Moreover, the interplay between $dN_{\rm PISN}/dM_{\rm SFR}(Z)$ and the adopted $Z$-evolution prescription can greatly enhance the differences with our results. Remarkably, the PISN rate obtained by \citealt{Stevenson_2019} extends up to three orders of magnitude downwards our pessimistic case (variation $P$ in Figure \ref{fig:var_stellar}). Indeed, on one hand their $dN_{\rm PISN}/dM_{\rm SFR}(Z)$ is between one and two orders of magnitude lower than in our variation $P$, at metallicities crucial for the PISN rate, close to $Z_{\rm max}\sim\:2\times 10^{-3}$ (compare our Figure \ref{fig:pisnpermass} with their Figure 2). This is also due to the fact that they consider an IMF upper limit of 150 $M_{\odot}$. Secondly, their galaxy metallicity distribution declines much more steeply in redshift with respect to ours. As explained in Sections \ref{sec:results_stellar} and \ref{sec:gal_var} for our results, this can dramatically lower the PISN rate, besides favouring the contribution from higher redshifts. Interestingly, this is the same reason why e.g. in \citealt{Tanikawa_2022} the PISN rate turns out to peak at much higher redshifts than ours, around $z\sim 6-7$. As already discussed, we obtain the same effect, albeit of smaller entity, by adopting a GSMF low-mass end slope varying with redshift, $\alpha_{\rm SMF}=\alpha_{\rm SMF}(z)$, producing a steeper decrease of the galaxy metallicity with redshift, and causing the PISN rate to peak up to $z\sim 4-5$ (see Section \ref{sec:gal_var}). If we change the \citealt{Curti_2020} FMR prescription adopted here to that of \citealt{Mannucci_2010}, we are even able to move the peak to $z>6$, as discussed below (Section \ref{sec:FMR_var}). Therefore, a redshift dependence of $\alpha_{\rm SMF}$ might motivate the employment of steeper $Z$-evolution prescriptions.

Noticeably, while like us most of previous works focus on $Z\geq 10^{-5}-10^{-4}$, \citealt{Tanikawa_2022,Tanikawa_2024} also consider very-low-metallicity, Pop III stars. As a result, e.g. in \citealt{Tanikawa_2024} the PISN rate features two distinct peaks, one around $z=2-3$ due to Pop II/I stars, and the other at $z\gtrsim 12$ due to Pop III. Depending on the considered variations, including different IMF upper limits, \citealt{Tanikawa_2022,Tanikawa_2024} find these two peak contributions to be roughly compatible, or the Pop III one to be dominant. We further discuss the contribution to the PISN rate from Pop III stars below.

\subsection{PI/CC ratio}

In Section \ref{sec:PI_CC_ratio}, we computed the ratio between PISN and CCSN rate, for our $Z_{\rm max}-\sigma_{\rm Z}$ combinations (see Table \ref{tab:PI_CC}). Our range of results exhibits an upper limit $\gtrsim 10^{-2}$, which is close to the value typically considered in the literature, coming from the simple integration of the IMF over the expected PISN and CCSN progenitor mass ranges. Adopting our Kroupa IMF, and integrating over the typical [$140-260$] $M_{\rm \odot}$ range for PISN, and [$8-50$] $M_{\rm \odot}$ for CCSN, gives a value of $\sim 1.5\times 10^{-2}$ for the PI/CC ratio. Considering the hundreds of observed CCSNe, and neglecting the different observational biases linked to these transients for the sake of simplicity, this would lead to expect at least 1 PISN detection so far, which is not the case. On the other hand, the lowest part of our range would tell us that PISNe are simply too rare with respect to CCSNe, making it trivial to understand why we never observed one. We stress that the differences in PISN and CCSN observation can be crucial and must be taken into account in order to make more robust statements. 

Using the missed PISN observation as a constraint, it is possible to infer an upper limit on the PI/CC ratio, as done in \citealt{Nicholl_2013}. They obtain a PI/CC ratio $<6\times 10^{-6}$ within $z<0.6$ (at 3$\sigma$), which is compatible with the lowest part of our range. Studies of this kind can be extremely valuable in constraining the parameters of stellar and galaxy evolution models, by excluding all parameter values that produce too high PI/CC ratios. However, these studies would inevitably suffer from important degeneracies. We leave a more accurate assessment regarding these matters to the follow-up work.

\subsection{PISNe in binaries}
As already outlined, so far we only considered PISNe arising from single stars, under the assumption that interactions in binaries (e.g. mass transfer) do not affect significantly the rate of PISN occurrence. In order to test this assumption, we run the most recent version of the \texttt{SEVN} code (\citealt{Iorio_2022}) to generate a synthetic population of binary stars. We then generate a population of single stars, and compute the $dN_{\rm PISN}/dM_{\rm SFR}$ for both populations. By assuming that a fraction $f_{\rm bin}=0.5$ of stars lie in binaries, we are able to compute the fraction of PISNe arising in binaries with respect to single stars, $f_{\rm bin}^{\rm PISN}$. 

We get $f_{\rm bin}^{\rm PISN}=0.56$, meaning that PISN events are distributed roughly equally among single stars and binaries. Indeed, we find that stellar mergers are the dominant process varying the masses of stars, while other processes such as common envelope and Roche-lobe overflow only play a secondary role. The number of stars entering and exiting the PISN mass range, by merging with another star, roughly balance each other. $f_{\rm bin}$ is slightly in favour of binaries due to our bottom-heavy IMF, making it so that stars with lower mass which merge into the PISN range are more abundant than higher mass stars merging out of that range.

We stress that this computation requires an assumption on $f_{\rm bin}$, which is still quite uncertain (e.g. \citealt{Sana_2012}). Nonetheless, we regard $f_{\rm bin}=0.5$ as a fiducial value, based on what established so far in the literature.

\subsection{FMR variation}\label{sec:FMR_var}
In this work, we adopted the FMR from \citealt{Curti_2020}, one of the most recent determinations (see Section \ref{sec:gal_model}). As we saw, this prescription leads to a drop in the $Z$-dependent SFRD at $z\lesssim 1$, at metallicities crucial for PISN production (Figure \ref{fig:CSFRD_in_Z}), causing the PISN rate to drop at $z\lesssim 1$ (e.g. Figure \ref{fig:Zmax_sigmaZ}). It is useful to consider an alternative FMR, in order to see how dependent our results are on this relation. We choose to follow \citealt{Mannucci_2010}, who present the following fit:
\begin{equation}
12+\log(O/H)=8.90+0.37m-0.14s-0.19m^2+0.12ms-0.054s^2,
\end{equation}
where $m=\log(M_{\rm \star})-10$ and $s=\log(SFR)$. Here, $\log Z=12+\log(O/H)-10.58$, as in \citealt{Boco_2021}. In the following, we will address \citealt{Curti_2020} as C20, and \citealt{Mannucci_2010} as M10.

\begin{figure*}
    \includegraphics[width=0.8\textwidth]{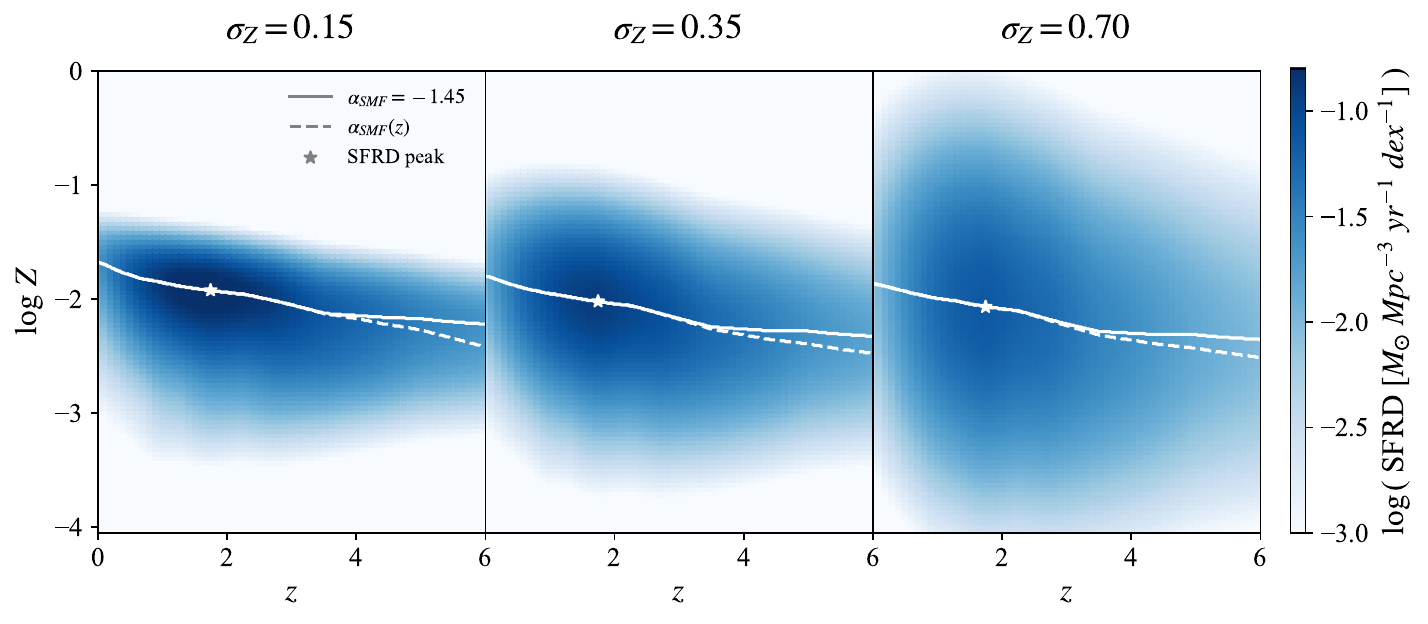}
    \caption{As Figure \ref{fig:CSFRD_in_Z}, for the \citealt{Mannucci_2010} FMR.}
    \label{fig:CSFRD_in_Z_M10}
\end{figure*}

\begin{figure}
        \centering
	\includegraphics[width=\columnwidth]{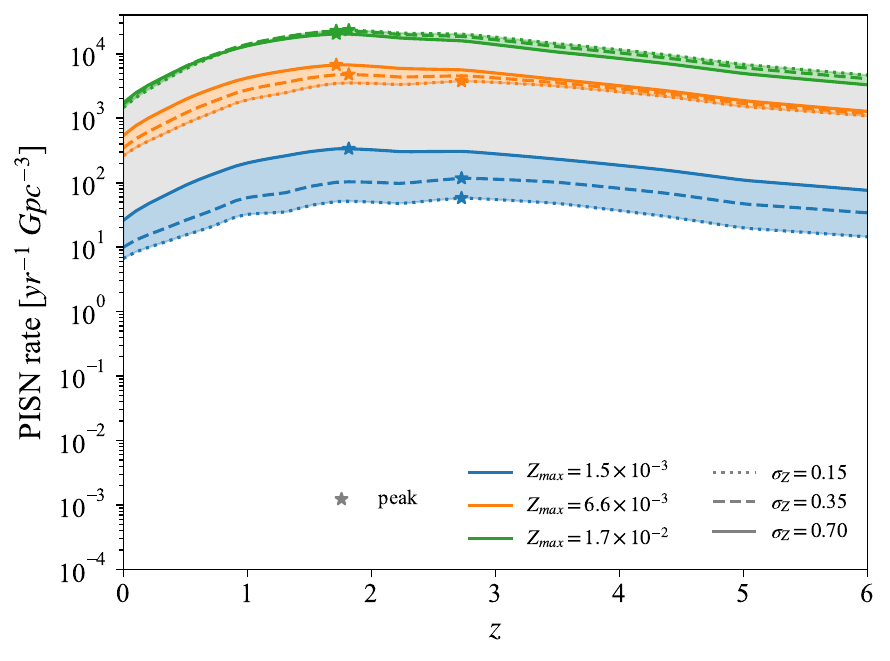}
    \caption{As Figure \ref{fig:Zmax_sigmaZ}, for the \citealt{Mannucci_2010} FMR.}
    \label{fig:rate_M10}
\end{figure}

\begin{figure*}
	\includegraphics[scale=0.5]{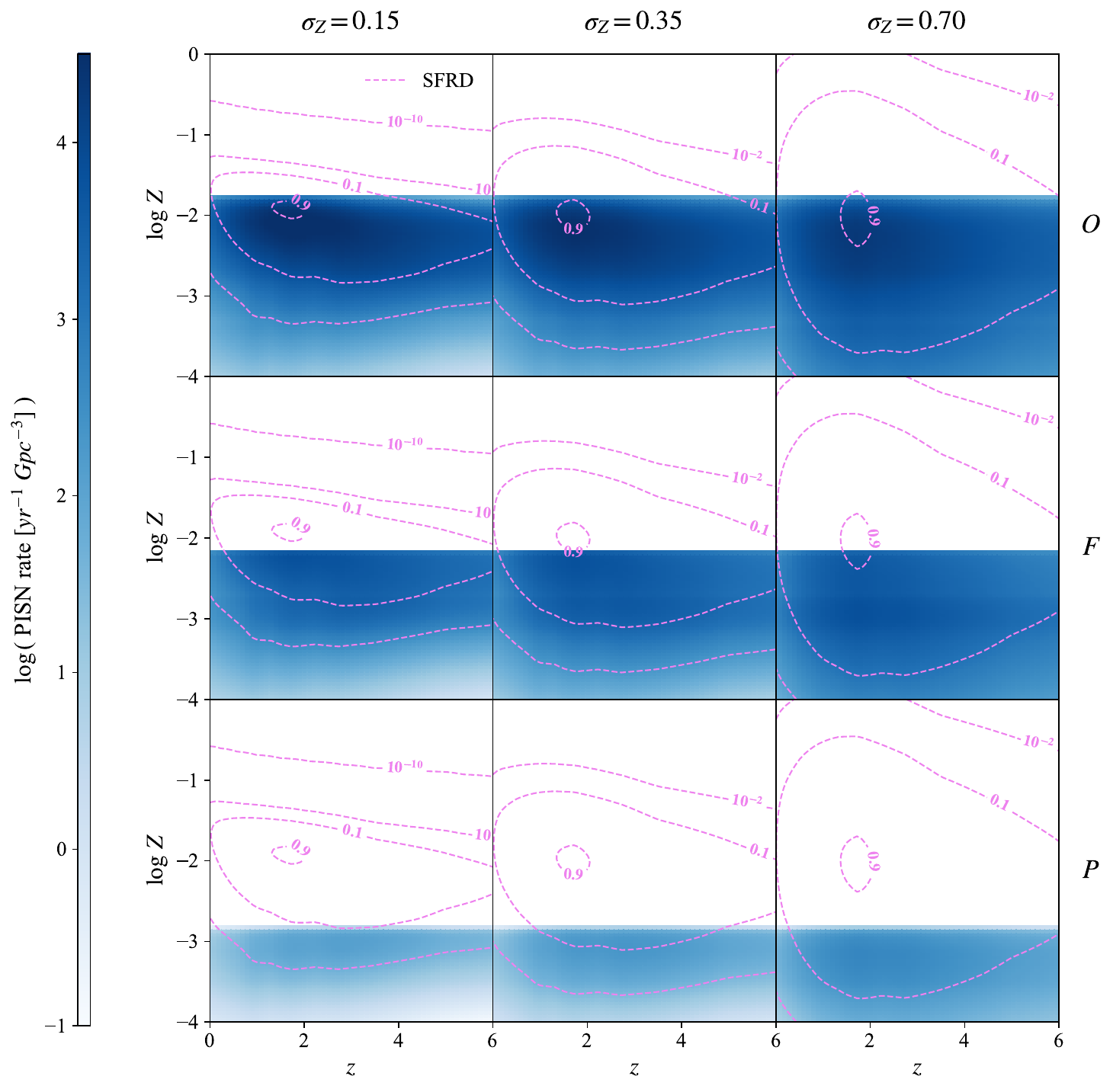}
    \caption{As Figure \ref{fig:rate_Z_z}, for the \citealt{Mannucci_2010} FMR.}
    \label{fig:rate_Z_z_M10}
\end{figure*}

\begin{figure}
        \centering
	\includegraphics[width=\columnwidth]{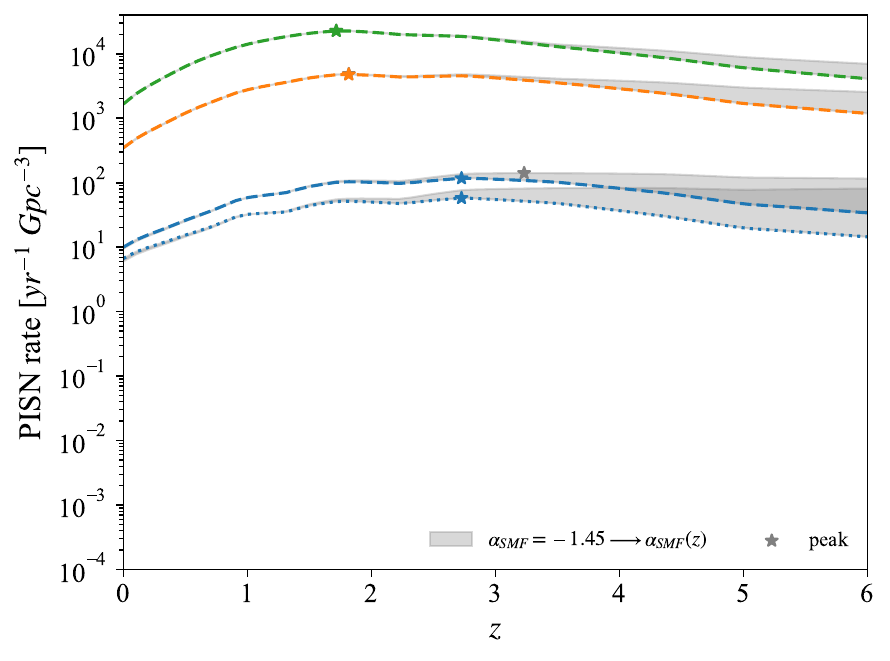}
    \caption{As in Figure \ref{fig:alpha_SMF}, for the \citealt{Mannucci_2010} FMR.}
    \label{fig:alpha_SMF_M10}
\end{figure}

Figure \ref{fig:CSFRD_in_Z_M10} shows the $Z$-dependent SFRD obtained by adopting the M10 FMR, for $\sigma_{\rm Z}=[0.15,0.35,0.70]$ and $\alpha_{\rm SMF}=-1.45$, to be compared with Figure \ref{fig:CSFRD_in_Z} for the C20 one. We show in Figure \ref{fig:rate_M10} the PISN rate we obtain for the same $Z_{\rm max}-\sigma_{\rm Z}$ combinations as Figure \ref{fig:Zmax_sigmaZ}. One can clearly see how this new FMR strongly reduces our range of results, which now sums up to just two/three orders of magnitude, as opposed to the five/seven orders of magnitude of Figure \ref{fig:Zmax_sigmaZ}. This is due to the fact that, for $\sigma_{\rm Z}=0.15$ and 0.35, the SFRD computed with the M10 FMR extends to lower metallicities, past $Z=10^{-3}$, with respect to the one computed with the C20 FMR. This is due to the steeper dependence of galaxy metallicity with stellar mass found in M10 (see also Figure 3 in C20). As a consequence, the PISN rate contribution coming from those metallicities is enhanced, as can be seen in Figure \ref{fig:rate_Z_z_M10}. Since this is where $Z_{\rm max}$ lies for our most pessimistic variations (Figure \ref{fig:pisnpermass}), the corresponding PISN rate is especially enhanced. Remarkably, in the case with $Z_{\rm max}=1.5\times 10^{-3}$ and $\sigma_{\rm Z}=0.15$ (lowest curve in Figures \ref{fig:Zmax_sigmaZ} and \ref{fig:rate_M10}), the difference in rate due to the FMR variation amounts to two/three orders of magnitude, reaching more than four orders of magnitude at $z=0$. On the other hand, in the case with $\sigma_{\rm Z}=0.70$ the PISN rate does not vary significantly since, as can be seen in the right panels of Figures \ref{fig:CSFRD_in_Z} and \ref{fig:CSFRD_in_Z_M10}, the SFRD distribution is quite similar in the two FMR variations. Moreover, the $\alpha_{\rm SMF}=\alpha_{\rm SMF}(z)$ variation produces an even steeper decrease of galaxy metallicity with redshift, with respect to the \cite{Curti_2020} FMR case. As a consequence, the PISN rate peak is brought to even higher redshifts, past $z=6$, as shown in Figure \ref{fig:alpha_SMF_M10}.

This warns us about how dependent the PISN rate can be on the adopted FMR. Our results must thus be taken with caution. We stress how it is the lowest end of this range to be affected, while its upper end appears to be immune. It again becomes clear how crucial the interplay between $Z_{\rm max}$ and $\sigma_{\rm Z}$ is in determining the PISN rate, it being the cause of these delicate dependencies.

\subsection{PISNe from Pop III stars}\label{sec:pisn_popIII}
It is traditionally believed that only low-$Z$ stars explode as PISNe. For this reason, due to their extremely low metallicities $Z\lesssim 10^{-10}$ (e.g. \citealt{Cassisi_1993}), very massive Population III stars are usually considered as main PISN progenitors (e.g. \citealt{Ober_1983,El_Eid_1983,Baraffe_2001,Fryer_2001,Umeda_2002,Heger_2002,Scannapieco_2005,Wise_2005,Langer_2007,Kasen_2011,Pan_2012,Dessart_2012,Yoon_2012,de_Souza_2013,de_Souza_2014,Whalen_2013,Whalen_2014,Smidt_2015,Magg_2016,Venditti_2023,Bovill_2024,Wiggins_2024}). On the other hand, our stellar evolution tracks and $M_{\rm CO}$ variations allow for PISNe up to $\sim$ solar metallicity, $\sim 1.7\times 10^{-2}$ for variation $O$. Combining with the SFRD, we find that the main PISN rate contribution comes from $Z\sim 10^{-3}-10^{-2}$, typical of Pop II/I stars. This result identifies Pop II/I stars as prominent PISN candidates. However, we stress that in this work we only consider $Z\geq 1\times 10^{-4}$ and $z\leq 6$. In order to study the contribution to the PISN rate coming from Pop III stars, and make a comparison with that from Pop II/I, one would need to extend the treatment to lower metallicities and higher redshifts. This task is hindered by the uncertainties surrounding relevant quantities like the Pop III SFRD and IMF, due to the lack of Pop III observations and the overall challenges of searching those very high redshifts. Nonetheless, we attempt at estimating the PISN rate contribution from Pop III stars, and compare it with our results for Pop II/I.\\

We compute the Pop III PISN rate analogously as in Equation \ref{eq:rate_z}. In particular, we consider the $Z=10^{-11}$ \texttt{PARSEC} stellar evolution tracks, introduced in \citealt{Costa_2023} inside the framework of \texttt{SEVN}. For the Pop III SFRD, we consider the determination by \citealt{Hartwig_2022} (see also \citealt{Santoliquido_2023}), obtained with the semi-analitical model A-SLOTH. Pop III stars are believed to follow a more top-heavy IMF with respect to Pop II/I, that might extend up to 1000 $M_{\odot}$ (e.g. \citealt{Hirano_2015}). This motivates us to adopt a Larson IMF (\citealt{Larson_1998}) of the form $\propto M^{\xi}\:e^{-M_c/M}$ (where $M\equiv M_{\rm ZAMS}$), defined up to 1000 $M_{\odot}$, with Kroupa-like slope $\xi=-2.3$, and characteristic mass $M_c=30\:M_{\odot}$. We note that varying the upper mass limit from 1000 to 300 $M_{\odot}$ does not affect the results, since the $Z=10^{-11}$ tracks we use predict a PISN progenitor mass range of $\sim[107-223]\:M_{\odot}$. Finally, we vary $M_{\rm CO}$ criterion for PISN as done in the rest of the work.\\

\begin{figure}
        \centering
	\includegraphics[width=\columnwidth]{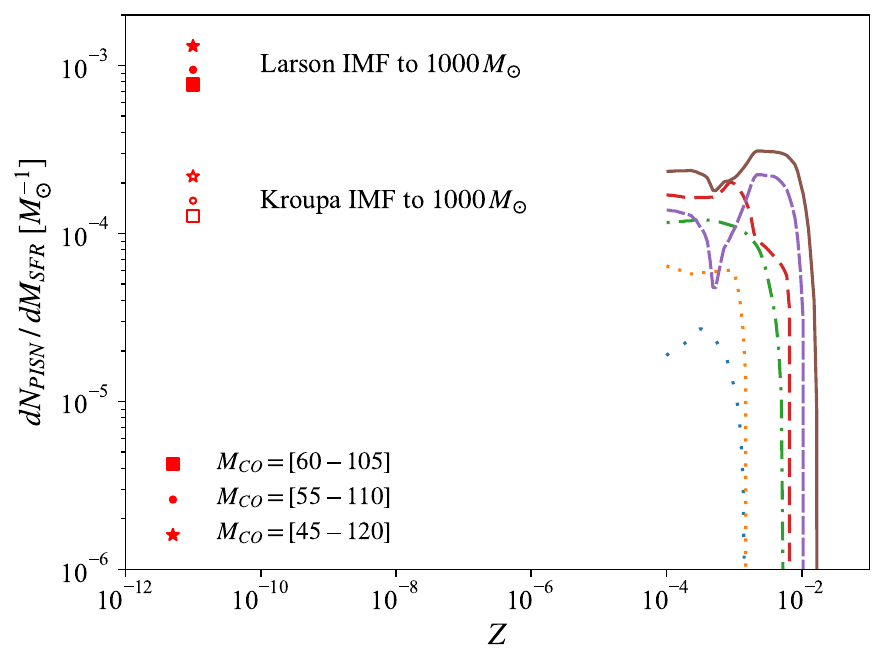}
    \caption{Number of PISNe per unit star forming mass produced by Pop III stars with $Z=10^{-11}$, adopting a Larson IMF up to 1000 $M_{\odot}$ (filled markers). We also show the case of a Kroupa IMF defined up to 1000 $M_{\odot}$ (empty markers). We indicate the $M_{\rm CO}$ variations with different shapes. For comparison, on the right is shown the same plot as in Figure \ref{fig:pisnpermass}.}
    \label{fig:PopIII_stellar_part}
\end{figure}

\begin{figure}
        \centering
	\includegraphics[width=\columnwidth]{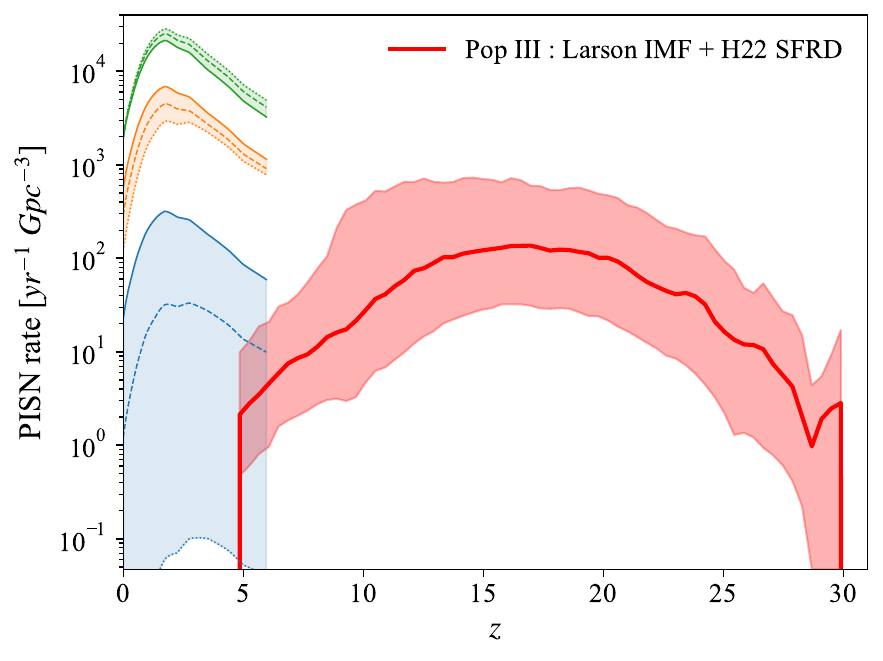}
    \caption{PISN rate from Pop III stars as a function of redshift, computed by adopting a Larson IMF and the Pop III SFRD determination by \citealt{Hartwig_2022}. The red band indicates the uncertainty coming from the 95\% credible interval of the \citealt{Hartwig_2022} SFRD, as computed in \citealt{Santoliquido_2023}. For clarity, we fix the fiducial, $M_{\rm CO}=[55-110]\:M_{\odot}$ variation. On the left we report the results obtained for Pop II/I stars, as in Figure \ref{fig:Zmax_sigmaZ}, for comparison.}
    \label{fig:PopIII_pisn_rate}
\end{figure}

In Figure \ref{fig:PopIII_stellar_part} we show the $dN_{\rm PISN}/dM_{\rm SFR}$ obtained with our Larson IMF, compared to the case of a Kroupa IMF defined up to 1000 $M_{\odot}$. As one can see, the Kroupa IMF produces a Pop III $dN_{\rm PISN}/dM_{\rm SFR}$ comparable to the Pop II/I one for the most optimistic stellar variations. On the contrary, the more top-heavy, Larson IMF elevates this quantity by almost an order of magnitude. Indeed, the Larson IMF increases the relative abundance of stars with masses in the PISN range, compared to the Kroupa IMF. Since this result depends only slightly on the chosen $M_{\rm CO}$ criterion, we fix it to the fiducial case in computing the PISN rate, that we show in Figure \ref{fig:PopIII_pisn_rate} for the Larson IMF. As one can see, the Pop III PISN rate lies around the same values of the most pessimistic results obtained for Pop II/I. Indeed, even though the \citealt{Hartwig_2022} Pop III SFRD is orders of magnitude lower than the one employed in this work for Pop II/I stars, the Larson IMF produces more PISNe per unit mass compared to the Kroupa IMF (Figure \ref{fig:PopIII_stellar_part}). Moreover, the inclusion of the whole SFRD contribution in $Z=10^{-11}$, allows to avoid the metallicity cut due to $Z_{\rm max}$ which, as discussed in depth in the paper, can lower significantly the PISN rate, especially in the lowest $\sigma_{\rm Z}$ cases. 

Previous Pop III PISN rate determinations in the literature tend to distribute in the middle/upper range of our results for Pop II/I stars (Figure \ref{fig:Zmax_sigmaZ}), with some even reaching our most optimistic results (\citealt{Scannapieco_2005,Wise_2005,Pan_2012,de_Souza_2013,de_Souza_2014,Magg_2016,Regos_2020,Tanikawa_2022,Venditti_2023,Wiggins_2024,Tanikawa_2024}). All in all, despite the uncertainties on the Pop III SFRD and IMF, these considerations suggest that Pop III could somewhat contribute to the PISN rate at $z\gtrsim 6$.

\section{Conclusions}\label{sec:conclusions}

In this work, we compute the PISN rate as a function of redshift up to $z=6$, by combining stellar evolution tracks with a semi-empirical determination of the metallicity-dependent cosmic star formation history. The aim is to study how the uncertainties in both stellar and galaxy evolution theory affect our knowledge about the rate of PISN occurrence throughout cosmic history. We vary stellar evolution code, $M_{\rm \rm CO}$ criterion to have PISN, and IMF upper limit, as well as the dispersion of the galaxy metallicity distribution and the low-mass end slope of the GSMFs. We find these variations to have a huge effect on our results. The PISN rate spans roughly three orders of magnitude under stellar variations. Their effect can be boiled down to the dependence on the maximum metallicity to have PISN, $Z_{\rm \rm max}$, and the height of the $dN_{\rm \rm PISN}/dM_{\rm \rm SFR}$ curve. Remarkably, the interplay between $Z_{\rm \rm max}$ and $\sigma_{\rm \rm Z}$ can extend the PISN rate range up to $\sim$ seven orders of magnitude, depending on redshift. The local, $z=0$ rate ranges from $4\times 10^{-4}$ to $2\times 10^3\:yr^{-1}\:Gpc^{-3}$, while the values at peak range from $\sim 10^{-1}$ to $3\times 10^4\:yr^{-1}\:Gpc^{-3}$. Moreover, prescribing a redshift dependence of $\alpha_{\rm \rm SMF}$ increases the PISN rate at $z>3-4$. Also the position of the peak gets affected, shifting from $z\lesssim 2$ up to $z\sim 4-5$ throughout all variations. This warns us about the delicate link between these two metallicity parameters, in choosing stellar evolution prescriptions and IMF leading to a certain $Z_{\rm \rm max}$, and in adopting recipes for the cosmic metallicity distribution of galaxies.

All in all, our limited knowledge makes the PISN rate very uncertain. On the other hand, the strong dependencies we find offer the chance to constrain these parameters based on possible PISN observations in the future, or the lack of PISN observations in the eventuality that these transients are never discovered. Even accounting for the significant degeneracies, this would help dispel the uncertainties around massive star evolution, such as the criterion for a star to undergo pair instability and the maximum stellar mass, and the evolution of star formation and galaxy metallicity throughout cosmic history. We will delve into these aspects in a follow-up work, dedicated to studying PISN detection with telescopes such as JWST, in which we will attempt at answering the question why PISN explosions have never been observed. 

Our fiducial model indicates galaxies with stellar masses between $\sim 10^9$ and $10^{10}$ $M_{\rm \rm \odot}$, and metallicities $\lesssim 10^{-2}$, to be the favourable PISN hosts, providing the main contribution to the PISN rate. These values shift to $M_{\rm \star}\sim 10^8-10^9\:M_{\rm \odot}$ and $Z\gtrsim 10^{-3}$ in our most pessimistic variation, and to $M_{\rm \star}\sim 10^{10}\:M_{\rm \odot}$ and $Z\sim 10^{-2}$ in our most optimistic one. This can represent useful information for the design of current and future campaigns aimed at observing these elusive transients, with facilities such as JWST, the \textit{Vera Rubin Observatory}, \textit{Euclid}, the \textit{Nancy Grace Roman Space Telescope}, the \textit{Zwicky Transient Facility}, and \textit{ULTIMATE-Subaru} (e.g. \citealt{Weinmann_2005,Whalen_2012,Smidt_2015,Kozyreva_2014,Hartwig_2018,Regos_2020,Moriya_2019,Moriya_2022,Moriya_2022_1,Tanikawa_2022,Tanikawa_2024}). 

Finally, our stellar evolution tracks and $M_{\rm CO}$ variations allow stars to explode as PISNe up to $\sim$ solar metallicity. As a consequence, we find peak metallicities for the PISN rate that suggest Pop II/I stars to be prominent PISN progenitors. This goes against the traditional expectation that PISNe come exclusively or preferentially from very-low-metallicity, Pop III stars, a point which is worth exploring in the future.

\section*{Acknowledgements}
The authors thank the anonymous referee for constructive comments, and Max Briel and David Hendriks for helpful discussions. This work was partially funded from the projects: ``Data Science methods for MultiMessenger Astrophysics \& Multi-Survey Cosmology'' funded by the Italian Ministry of University and Research, Programmazione triennale 2021/2023 (DM n.2503 dd. 9 December 2019), Programma Congiunto Scuole; EU H2020-MSCA-ITN-2019 n. 860744 \textit{BiD4BESt: Big Data applications for black hole Evolution STudies}; Italian Research Center on High Performance Computing Big Data and Quantum Computing (ICSC), project funded by European Union - NextGenerationEU - and National Recovery and Resilience Plan (NRRP) - Mission 4 Component 2 within the activities of Spoke 3 (Astrophysics and Cosmos Observations);  European Union - NextGenerationEU under the PRIN MUR 2022 project n. 20224JR28W "Charting unexplored avenues in Dark Matter"; INAF Large Grant 2022 funding scheme with the project "MeerKAT and LOFAR Team up: a Unique Radio Window on Galaxy/AGN co-Evolution; INAF GO-GTO Normal 2023 funding scheme with the project "Serendipitous H-ATLAS-fields Observations of Radio Extragalactic Sources (SHORES)". The authors declare no conflict of interest.

\appendix

\section{He core mass criterion}\label{appendix:He}
In this Section, we explore how adopting the criterion on $M_{\rm He}$, instead of $M_{\rm CO}$, affects our results. We impose that only stars with $M_{\rm He}$ in the range [64-135] $M_{\rm \odot}$ explode as PISNe, as adopted in \texttt{SEVN} (\citealt{Spera_2017}). We consider [70-120] and [60-140] $M_{\rm \odot}$ as pessimistic and optimistic ranges respectively, based on previous works (see references in Section \ref{sec:st_ev_model}). Table \ref{tab:stellar_combos_He} shows the stellar variations we considered, namely the same as in Table \ref{tab:stellar_combos} but with $M_{\rm He}$ ranges. Note that $Z_{\rm max}$ is modified as a consequence.

\begin{table}
\centering
\captionof{table}{As in Table \ref{tab:stellar_combos}, with $M_{\rm He}$ mass ranges and, consequently, different $Z_{\rm max}$.}
\begin{tabular}{|c|c|c|c|c|}
 \hline
 name & stellar code & $M_{\rm He}/M_{\rm \odot}$ & $M_{\rm up}/M_{\rm \odot}$ & $Z_{\rm max}$ \\
 \hline
 P & \texttt{FRANEC} & 70-120 & 150 & 1.5$\times 10^{-3}$ \\
 \hline
 M1 & \texttt{PARSEC-I} & 64-135 & 150 & 2.0$\times 10^{-3}$ \\
 \hline
 M2 & \texttt{FRANEC} & 60-140 & 150 & 3.1$\times 10^{-3}$ \\
 \hline
 F & \texttt{PARSEC-I} & 64-135 & 300 & 7.9$\times 10^{-3}$ \\
 \hline
 M3 & \texttt{PARSEC-II} & 60-140 & 150 & 8.7$\times 10^{-3}$ \\
 \hline
 O & \texttt{PARSEC-II} & 60-140 & 300 & 1.5$\times 10^{-2}$ \\
 \hline
 \label{tab:stellar_combos_He}
\end{tabular}
\end{table}

\begin{figure}
        \centering
	\includegraphics[width=\columnwidth]{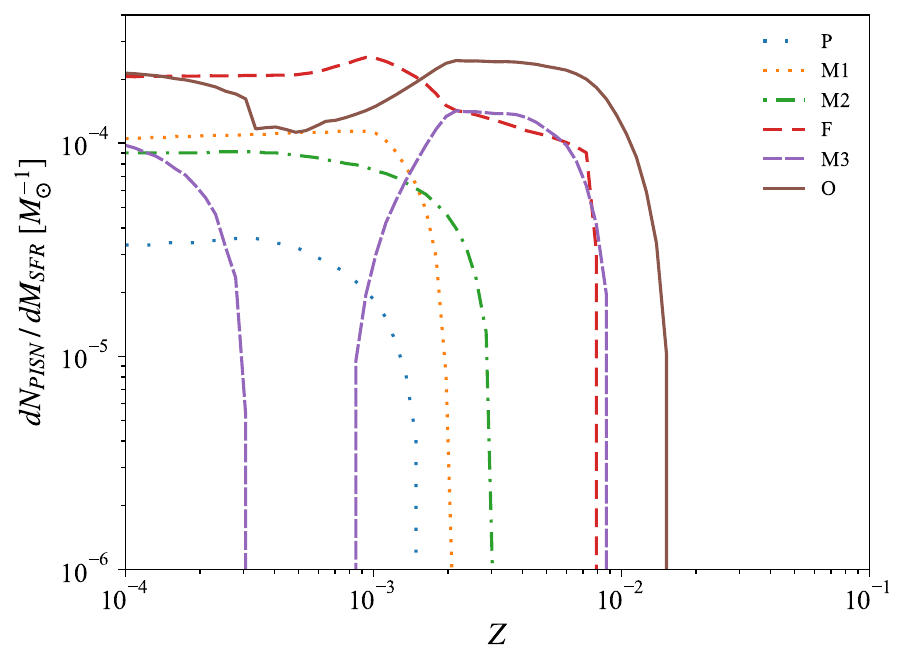}
    \caption{As in Figure \ref{fig:pisnpermass}, using the $M_{\rm He}$ criterion (see Table \ref{tab:stellar_combos_He}).}
    \label{fig:pisnpermass_He}
\end{figure}

As shown in Figure \ref{fig:pisnpermass_He}, the shape of $dN_{\rm PISN}/dM_{\rm SFR}$ is similar to the $M_{\rm CO}$ case (Figure \ref{fig:pisnpermass}), except for variation $M3$, which exhibits a harsh drop at metallicities between $\sim 4$ and $9\times 10^{-3}$. This is due to envelope undershooting, reducing the mass of the core and thus preventing stars in the corresponding mass range from going into PISN. The effect on $M_{\rm He}$ is stronger than on $M_{\rm CO}$, where it manifests as a softer drop. See \citealt{Costa_2021} for a detailed explanation of this process in \texttt{PARSEC}. We note that this drop is not particularly relevant for the PISN rate, the main contribution coming from the bulge between $Z\sim 2\times 10^{-3}$ and $7\times 10^{-2}$. $Z_{\rm max}$ is affected in a non-negligible way in most variations, as shown in Table \ref{tab:stellar_combos_He}. The values of $dN_{\rm PISN}/dM_{\rm SFR}$ change by some small factor. Note that we chose core mass ranges which are not correspondent to each other, between the $M_{\rm CO}$ and $M_{\rm He}$ cases, which is mostly causing these differences. 

\begin{figure}
        \centering
	\includegraphics[width=\columnwidth]{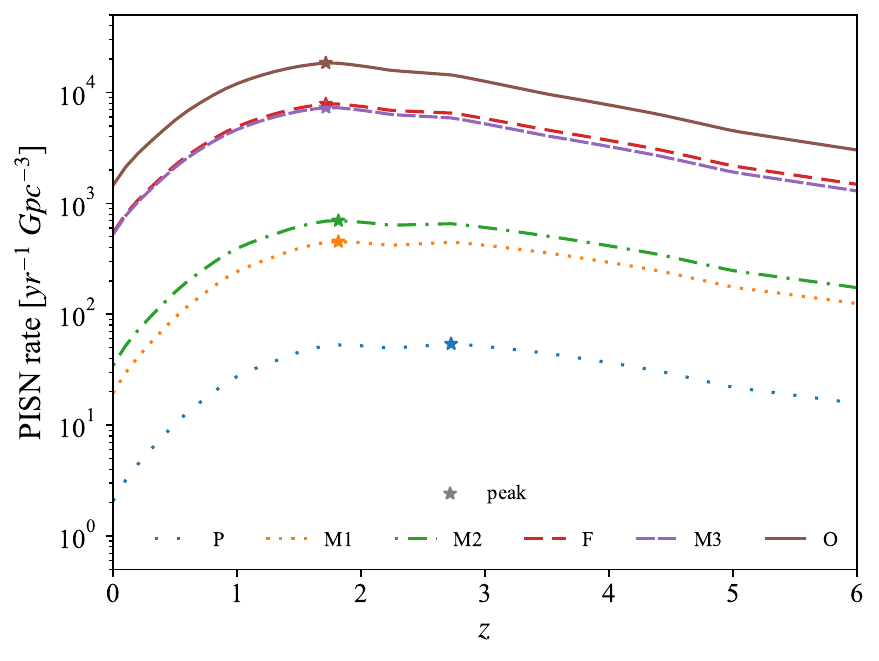}
    \caption{As in Figure \ref{fig:var_stellar}, using the $M_{\rm He}$ criterion.}
    \label{fig:var_stellar_He}
\end{figure}

In Figure \ref{fig:var_stellar_He}, we show the PISN rate resulting from each stellar variation. As one can see, the rates only change by some small factor, in any case less than one order of magnitude, mainly due to the different $Z_{\rm max}$ (compare with Figure \ref{fig:var_stellar}). In particular, variations $P$ and $O$ are not subject to dramatic changes, so that the range of results is left substantially unaffected.

\begin{figure}
        \centering
	\includegraphics[width=\columnwidth]{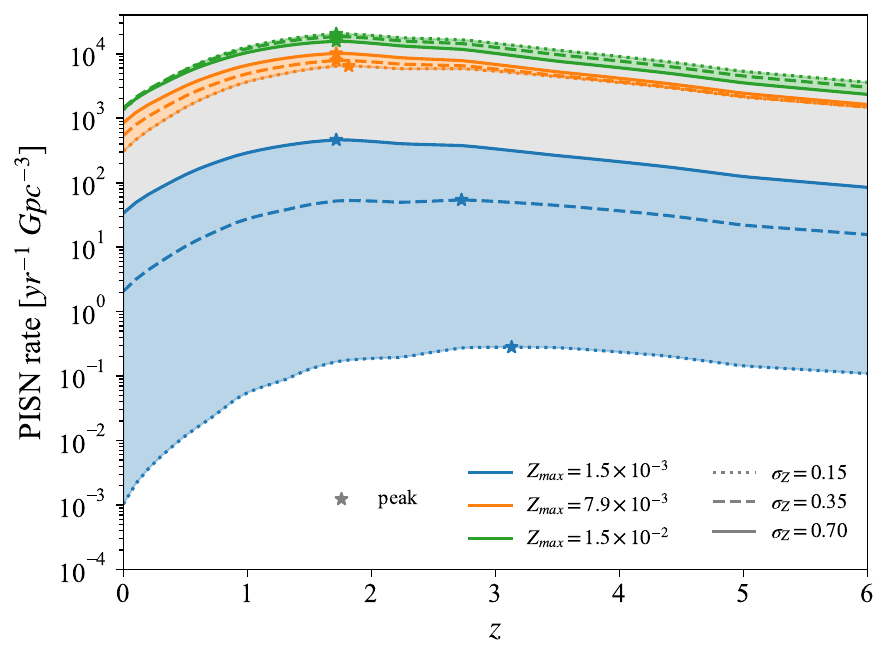}
    \caption{As in Figure \ref{fig:Zmax_sigmaZ}, using the $M_{\rm He}$ criterion.}
    \label{fig:Zmax_sigmaZ_He}
\end{figure}

\begin{figure}
        \centering
	\includegraphics[width=\columnwidth]{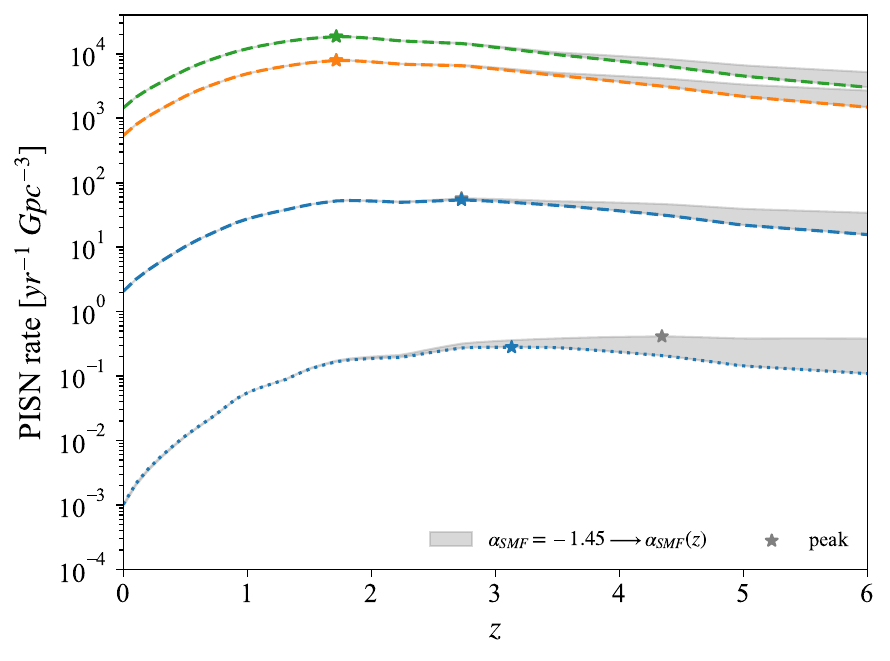}
    \caption{As in Figure \ref{fig:alpha_SMF}, using the $M_{\rm He}$ criterion.}
    \label{fig:alpha_SMF_He}
\end{figure}

Analogously, we find that also applying our galactic variations does not lead to significant differences, with the PISN rate ranges being only slightly shrinked with respect to the $M_{\rm CO}$ case, as shown in Figures \ref{fig:Zmax_sigmaZ_He} and \ref{fig:alpha_SMF_He}.

Overall, we find that varying between $M_{\rm CO}$ and $M_{\rm He}$ criterion does not change our results significantly.

\section{Additional IMF upper limit variation}\label{appendix:IMF600}
Throughout this work, for simplicity we assumed a Kroupa IMF, and explored variations on its upper limit, $M_{\rm up}=150$ and $300\:M_{\rm \odot}$. Extending this IMF to higher masses does not affect our results significantly. In order to prove this point, we consider an additional variation with $M_{\rm up}=600\:M_{\rm \odot}$.

Indeed, considering the typical $M_{\rm ZAMS}$ of $\sim [140,260]\:M_{\rm \odot}$ for PISN progenitors, one would not expect IMFs with $M_{\rm up}>260\:M_{\rm \odot}$ to have any effect. However, the stellar evolution codes we consider here produce $M_{\rm ZAMS}$ ranges varying significantly with metallicity, as shown in Table \ref{tab:mzams_intervals}. In particular, they predict even VMSs beyond $300\:M_{\rm \odot}$ to end their life as PISN, at high metallicity. Therefore, in principle extending the IMF to 600 $M_{\rm \odot}$ could play a role.

\begin{figure}
        \centering
	\includegraphics[width=\columnwidth]{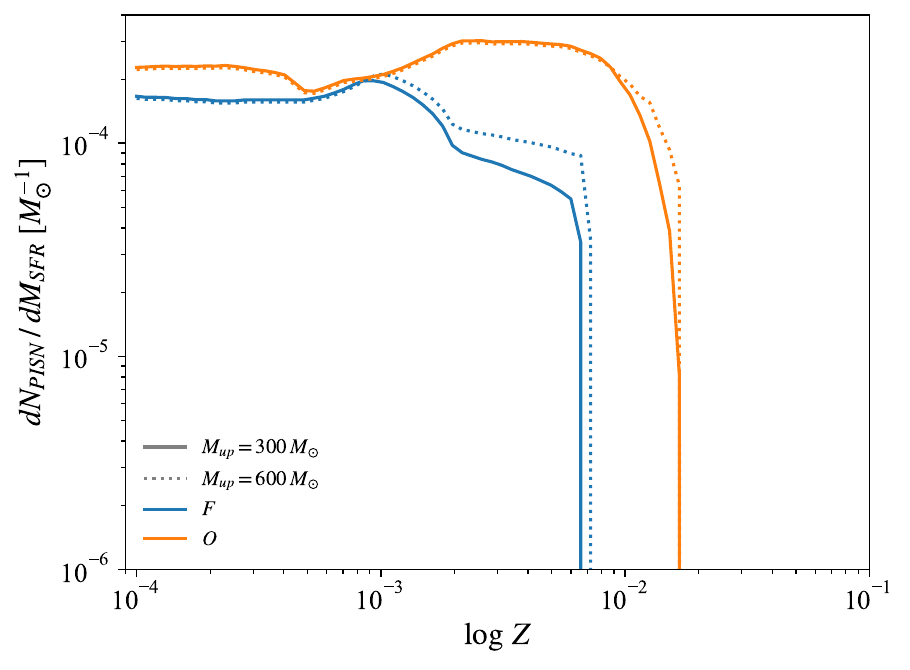}
    \caption{As in Figure \ref{fig:pisnpermass}, for $M_{\rm up}=300$ and $600\:M_{\rm \odot}$ (solid and dotted lines respectively), and for $M_{\rm CO}\in[60-105]$ and $[45,120]\:M_{\rm \odot}$ (blue and orange lines). The \texttt{PARSEC-II} stellar evolution tracks have been employed.}
    \label{fig:pisn_per_mass_IMF600}
\end{figure}

\begin{figure}
        \centering
	\includegraphics[width=\columnwidth]{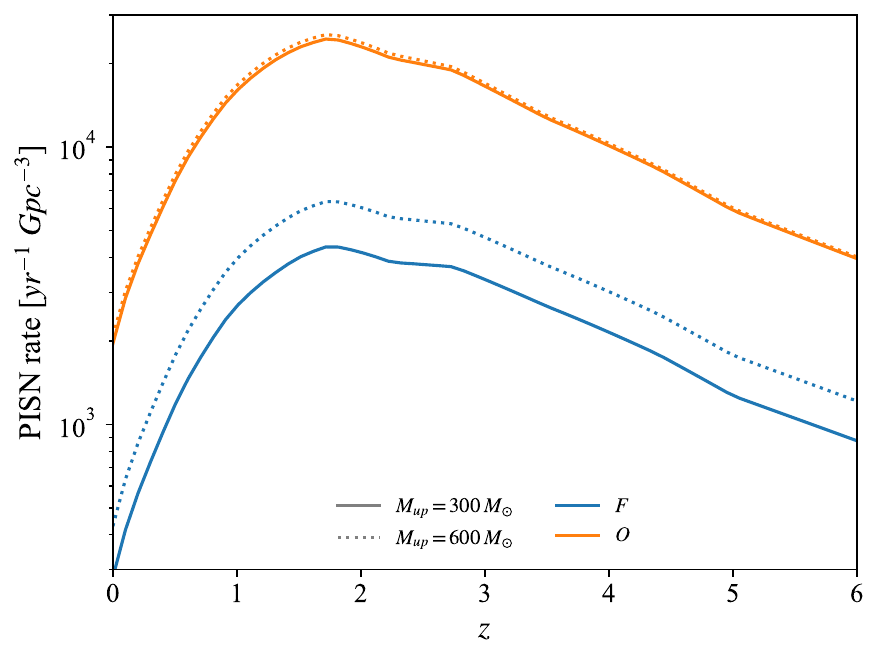}
    \caption{PISN rate as a function of redshift, obtained with the \texttt{PARSEC-II} tracks for the stellar variations shown in Figure \ref{fig:pisn_per_mass_IMF600}. Galactic parameters are fixed to $\sigma_{\rm Z}=0.35$ and $\alpha_{\rm SMF}=-1.45$.}
    \label{fig:rate_IMF600}
\end{figure}

Among our set of stellar variations, we select $F$ and $O$, since they have respectively the lowest and highest $Z_{\rm max}$ among all possible variations with $M_{\rm up}=300\:M_{\rm \odot}$. For simplicity, we do not consider different variations on $M_{\rm CO}$, which do not play a significant role in this case. Figure \ref{fig:pisn_per_mass_IMF600} shows the $dN_{\rm PISN}/dM_{\rm SFR}$ we obtain by imposing $M_{\rm up}=300$ and $600\:M_{\rm \odot}$. In variation $O$, setting $M_{\rm up}$ to 600 $M_{\rm \odot}$ has an appreciable effect only at $Z\gtrsim 10^{-2}$, where it slightly increases $dN_{\rm PISN}/dM_{\rm SFR}$. In variation $F$, $dN_{\rm PISN}/dM_{\rm SFR}$ gets increased starting from $Z\lesssim 10^{-3}$, and also $Z_{\rm max}$ is slightly higher. As one can see in Figure \ref{fig:rate_IMF600}, the effects on the PISN rate are negligible for variation $O$, while in variation $F$ the rate is increased only by a small factor, between 1 and 2. This is because in the first case $Z_{\rm max}$ is greater than the SFRD peak (see Section \ref{sec:gal_model}), therefore the main contribution to the PISN rate is already included. In variation $F$, $Z_{\rm max}$ is lower than the SFRD peak, so the rate is more sensible to this variation, even though it is affected only in a minor way.

All in all, we find that varying $M_{\rm up}$ from 300 to 600 $M_{\rm \odot}$ does not have a significant effect on the PISN rate. This is due to the fact that our Kroupa IMF prescribes a low number of stars with mass between 300 and 600 $M_{\rm \odot}$. Adopting a more top-heavy IMF might lead to different results.

\section*{Data Availability}
The data used in this work can be made available upon request.



\bibliographystyle{mnras}
\bibliography{PISN_first_draft} 






\bsp	
\label{lastpage}
\end{document}